\documentclass[twocolumn]{aastex62}
\usepackage{graphicx,ifthen,url,float,color,ulem,amsmath,ragged2e}

\newcommand {\sourcelong}{MM\,J100026.36$+$021527.9}
\newcommand {\source}{{\sc Mambo--9}}

\newcommand {\coa}{$^{12}$CO($J\!=$1$\rightarrow$0)}
\newcommand {\coe}{$^{12}$CO($J\!=$5$\rightarrow$4)}
\newcommand {\cof}{$^{12}$CO($J\!=$6$\rightarrow$5)}
\newcommand {\htwoo}{p-H$_{2}$O(2$_{1,1}\!\rightarrow$2$_{0,2}$)}
\newcommand {\kms}{km s$^{-1}$}

\def\ltsima{$\; \buildrel < \over \sim \;$}
\def\simlt{\lower.5ex\hbox{\ltsima}}
\def\gtsima{$\; \buildrel > \over \sim \;$}
\def\simgt{\lower.5ex\hbox{\gtsima}}
\newcommand {\uJy}{$\mu$Jy}
\newcommand {\um}{$\mu$m}

\newcommand{\msun}{{\rm\,M$_\odot$}}
\newcommand{\sfr}{{\rm\,M$_\odot$\,yr$^{-1}$}}
\newcommand{\lsun}{{\rm\,L$_\odot$}}

\newcommand{\lpeak}{$\lambda_{\rm peak}$}
\newcommand{\lir}{L$_{\rm IR}$}
\received{September 3, 2019}
\accepted{October 29, 2019}

\shorttitle{Characterization of a z=5.8 DSFG}
\shortauthors{Casey et al.}

\begin{document}

\title{\sc Physical characterization of an unlensed dusty star-forming galaxy at z\,=\,5.85}

\correspondingauthor{Caitlin M. Casey}
\email{cmcasey@utexas.edu}

\author[0000-0002-0930-6466]{Caitlin M. Casey}
\affil{Department of Astronomy, The University of Texas at Austin, 2515 Speedway Blvd Stop C1400, Austin, TX 78712, USA}

\author[0000-0002-7051-1100]{Jorge A. Zavala}
\affil{Department of Astronomy, The University of Texas at Austin, 2515 Speedway Blvd Stop C1400, Austin, TX 78712, USA}

\author{Manuel Aravena}
\affil{N\'{u}cleo de Astronomi\'{i}a, Facultad de Ingenier\'{i}a y Ciencias, Universidad Diego Portales, Av. Ej\'{e}rcito 441, Santiago, Chile}

\author{Matthieu B\'{e}thermin}
\affil{Aix-Marseille Universit\'{e}, CNRS, CNES, LAM, Marseille, France}

\author{Karina I. Caputi}
\affil{Kapteyn Astronomical Institute, University of Groningen, P.O. Box 800, 9700AV Groningen, The Netherlands}
\affil{Cosmic Dawn Center (DAWN)}

\author{Jaclyn B. Champagne}
\affil{Department of Astronomy, The University of Texas at Austin, 2515 Speedway Blvd Stop C1400, Austin, TX 78712, USA}

\author{David L. Clements}
\affil{Imperial College London, Blackett Laboratory, Prince Consort Road, London, SW7 2AZ, UK}

\author[0000-0001-9759-4797]{Elisabete da Cunha}
\affil{International Centre for Radio Astronomy Research, University of Western Australia, 35 Stirling Hwy, Crawley, WA 6009, Australia}
\affil{Research School of Astronomy and Astrophysics, The Australian National University, Canberra ACT 2611, Australia}
\affil{ARC Centre of Excellence for All Sky Astrophysics in 3 Dimensions (ASTRO 3D)}

\author[0000-0003-3627-7485]{Patrick Drew}
\affil{Department of Astronomy, The University of Texas at Austin, 2515 Speedway Blvd Stop C1400, Austin, TX 78712}

\author[0000-0001-8519-1130]{Steven L. Finkelstein}
\affil{Department of Astronomy, The University of Texas at Austin, 2515 Speedway Blvd Stop C1400, Austin, TX 78712, USA}

\author{Christopher C. Hayward}
\affil{Center for Computational Astrophysics, Flatiron Institute, 162 Fifth Avenue, New York, NY 10010, USA}

\author{Jeyhan S. Kartaltepe}
\affil{School of Physics and Astronomy, Rochester Institute of Technology, Rochester, NY 14623, USA}

\author{Kirsten Knudsen}
\affil{Department of Earth and Space Sciences, Chalmers University of Technology, Onsala Space Observatory, SE-43992 Onsala, Sweden}

\author{Anton M. Koekemoer}
\affiliation{Space Telescope Science Institute, 3700 San Martin Dr., Baltimore, MD 21218, USA}

\author{Georgios E. Magdis}
\affil{Cosmic Dawn Center (DAWN)}
\affil{Niels Bohr Institute, University of Copenhagen, Lyngbyvej 2, DK-2100 Copenhagen, Denmark}

\author{Allison Man}
\affil{Dunlap Institute for Astronomy \&\ Astrophysics, 50 St. George Street, Toronto, ON M5S 3H4, Canada}

\author{Sinclaire M. Manning}
\affil{Department of Astronomy, The University of Texas at Austin, 2515 Speedway Blvd Stop C1400, Austin, TX 78712, USA}

\author{Nick Z. Scoville}
\affil{California Institute of Technology, MC 249-17, 1200 East California Boulevard, Pasadena, CA 91125, USA}

\author{Kartik Sheth}
\affil{NASA Headquarters, 300 E Street SW, Washington, DC 20546, USA}

\author{Justin Spilker}
\affil{Department of Astronomy, The University of Texas at Austin, 2515 Speedway Blvd Stop C1400, Austin, TX 78712, USA}

\author{Johannes Staguhn}
\affil{The Henry A. Rowland Department of Physics and Astronomy, Johns Hopkins University, 3400 North Charles Street, Baltimore, MD 21218, USA}
\affil{Observational Cosmology Lab, Code 665, NASA Goddard Space Flight Center, Greenbelt, MD 20771, USA}

\author[0000-0003-4352-2063]{Margherita Talia}
\affil{Dipartimento di Fisica e Astronomia, Universit\`{a} di Bologna, Via Gobetti 93/2, I-40129, Bologna, Italy}

\author{Yoshiaki Taniguchi}
\affil{The Open University of Japan, 2-11 Wakaba, Mihama-ku, Chiba 261-8586, Japan}

\author{Sune Toft}
\affil{Cosmic Dawn Center (DAWN)}
\affil{Niels Bohr Institute, University of Copenhagen, Lyngbyvej 2, DK-2100 Copenhagen, Denmark}

\author{Ezequiel Treister}
\affil{Instituto de Astrof\'{i}sica and Centro de Astroingenier\'{i}a, Facultad de F\'{i}sica, Pontificia Universidad Cat\'{o}lica de Chile, Casilla 306, Santiago 22, Chile}

\author{Min Yun}
\affil{Department of Astronomy, University of Massachusetts Amherst, 710 N. Pleasant Street, Amherst, MA 01003, USA}

\begin{abstract}
We present a physical characterization of \sourcelong\
(a.k.a. ``\source''), a dusty star-forming galaxy (DSFG) at
$z=5.850\pm0.001$.  This is the highest redshift unlensed DSFG (and
fourth most distant overall) found to-date, and is the first source
identified in a new 2\,mm blank-field map in the COSMOS field. Though
identified in prior samples of DSFGs at 850\um--1.2\,mm with unknown
redshift, the detection at 2\,mm prompted further follow-up as it
indicated a much higher probability that the source was likely to sit
at $z>4$.  Deep observations from the Atacama Large Millimeter and
submillimeter Array (ALMA) presented here confirm the redshift through
the secure detection of \cof\ and \htwoo. \source\ is comprised of a
pair of galaxies separated by 6\,kpc with corresponding star-formation
rates of 590\,\sfr\ and 220\,\sfr, total molecular hydrogen gas mass
of (1.7$\pm$0.4)$\times10^{11}$\,\msun, dust mass of
(1.3$\pm$0.3)$\times10^{9}$\,\msun\ and stellar mass of
(3.2$^{+1.0}_{-1.5}$)$\times10^{9}$\,\msun.  The total halo mass,
(3.3$\pm$0.8)$\times10^{12}$\,\msun, is predicted to exceed
$>10^{15}$\,\msun\ by $z=0$.  The system is undergoing a merger-driven
starburst which will increase the stellar mass of the system tenfold
in $\tau_{\rm depl}=40-80$\,Myr, converting its large molecular gas
reservoir (gas fraction of 96$^{+1}_{-2}$\%) into stars.  \source\
evaded firm spectroscopic identification for a decade, following a
pattern that has emerged for some of the highest redshift DSFGs
found. And yet, the systematic identification of unlensed DSFGs like
\source\ is key to measuring the global contribution of
obscured star-formation to the star-formation rate density at
$z\simgt4$, the formation of the first massive galaxies, and the
formation of interstellar dust at early times ($\simlt$1\,Gyr).
\end{abstract}


\section{Introduction} \label{sec:intro}

The most extreme star-forming galaxies in the Universe pose unique
challenges for galaxy formation theory
\citep[e.g.][]{fardal01a,baugh05a,lacey08a,gonzalez11a,narayanan15a}.
Because dust is a byproduct of star-formation, the ubiquity of high
star-formation rate galaxies at $z\sim2$ means that dust-rich systems
were common and dominated the cosmic star-forming budget for several
billions of years \citep*[e.g.][]{madau14a,casey14a}.  However, the
identification of these Dusty Star-Forming Galaxies (DSFGs) out to
higher redshifts ($z\simgt4$), in the first two Gyr after the Big
Bang, has proven exceedingly difficult.  While extraordinary
discoveries of DSFGs exist out to $z\sim7$ \citep[SPT0311 being the
  highest-$z$ DSFG found to-date,][]{strandet17a,marrone18a}, their
total contribution to the cosmic star-formation budget is
unconstrained during this early epoch \citep{casey18a}.  Contradictory
results have been presented in the literature, with some claiming
that DSFGs play an insignificant role in $z>4$ star-formation with
less than 10\%\ of the total \citep[e.g.][]{koprowski17a}, while
others suggest DSFGs may dominate cosmic star-formation at a level
exceeding 90\%\ in the first Gyr \citep{rowan-robinson16a}.  Several
other works suggest the truth might lie between these two extremes
\citep[e.g.][]{bethermin17a,zavala18a}, though data to constrain this
epoch is sparse leaving estimates highly uncertain.

Identifying individual DSFGs at early epochs in the Universe's history
is critical to our understanding of how massive galaxies assemble and,
independently, how vast dust reservoirs are formed so early in a
galaxy's history, whether it be from asymptotic giant branch (AGB)
stars, supernovae or efficient ISM grain growth
\citep[e.g.][]{matsuura06a,matsuura09a,zhukovska08a,asano13a,jones13a,dwek14a}.

In this paper we describe the detection and characterization of the
highest-redshift, unlensed DSFG to-date, confirmed at $z=5.85$
\citep[see also][for an independent analysis of this source]{jin19a}.
This galaxy was identified as a submillimeter-luminous source by
MAMBO, AzTEC and SCUBA-2
\citep{bertoldi07a,aretxaga11a,casey13a,geach17b} though it lacked a
secure redshift identification for many years. The source was
identified independently by many groups as a high-redshift candidate,
and was recently spectroscopically observed with ALMA as presented in
\citet{jin19a}. We corroborate their proposed redshift solution
through independent ALMA observations in this paper.  Here we present
a multi-wavelength characterization of the source in order to
constrain its physical drivers and characteristics.
\S~\ref{sec:observations} presents our observations,
\S~\ref{sec:results} presents calculations of critical physical
quantities like dynamical, gas, stellar and dust mass, and
\S~\ref{sec:discussion} presents our interpretation of this galaxy's
physical drivers and broader context.  Throughout we assume a {\it
  Planck} cosmology \citep{planck-collaboration18a} and assume a
Chabrier initial mass function for the purpose of calculation stellar
masses and star-formation rates \citep{chabrier03a}.

\begin{table*}
\caption{\source\ Photometry}
\centering
\begin{tabular}{lccrrrl}
\hline\hline
Band & Wavelength & Units & Component A & Component B & Total (A+B) & Data Reference \\
\hline
{\it HST}-F606W & 606\,nm & nJy &  (3.7$\pm$8.8) & (--0.5$\pm$8.8) & (10.2$\pm$25.7) & \citet{koekemoer11a} \\
{\it HST}-F814W & 814\,nm & nJy & (9.2$\pm$11.5) & (--2.6$\pm$11.5) & (--3.2$\pm$31.7) & \citet{koekemoer11a} \\
{\it HST}-F125W & 1.25\,\um & nJy & (1.8$\pm$14.0) & (34.3$\pm$14.0) & (36.6$\pm$47.0) & \citet{koekemoer11a} \\
{\it HST}-F160W & 1.60\,\um & nJy & (5.3$\pm$13.8) & (15.4$\pm$13.8) & (50.0$\pm$41.0) & \citet{koekemoer11a} \\
IRAC-CH1 & 3.6\um\ & nJy & --- & --- & 87$\pm$29 & \citet{ashby15a} \\
IRAC-CH2 & 4.5\um\ & nJy & --- & --- & 186$\pm$37 & \citet{ashby15a} \\
MIPS24 & 24\um\    & \uJy & --- & --- & (10$\pm$18) & \citet{le-floch09a}\\
PACS & 100\um\   & \uJy & --- & --- & (48$\pm$152) & \citet{lutz11a} \\
PACS & 160\um\   & \uJy & --- & --- & (--56$\pm$276) & \citet{lutz11a} \\
SPIRE & 250\um\   & mJy & --- & --- & (2.9$\pm$5.8) & \citet{oliver12a} \\
SPIRE & 350\um\   & mJy & --- & --- & (2.9$\pm$6.3) & \citet{oliver12a} \\
SCUBA-2 & 450\um\ & mJy & --- & --- & (2.32$\pm$5.82) & \citet{casey13a} \\
SPIRE & 500\um\   & mJy & --- & --- & (4.9$\pm$6.8) & \citet{oliver12a} \\
SCUBA-2 & 850\um\ & mJy & --- & --- & 5.84$\pm$0.87 & \citet{geach17b} \\
ALMA-B7 & 871\um\  & mJy   & 4.032$\pm$0.048 & 1.486$\pm$0.280 & 5.908$\pm$0.052 & {\sc this work}\\
ALMA-B7 & 876\um\  & mJy   & 3.938$\pm$0.042 & 1.410$\pm$0.285 & 5.262$\pm$0.041 & {\sc this work}\\
ALMA-B7 & 902\um\  & mJy   & 3.851$\pm$0.050 & 1.180$\pm$0.246 & 5.220$\pm$0.049 & {\sc this work}\\
ALMA-B7 & 908\um\  & mJy   & 3.853$\pm$0.065 & 1.650$\pm$0.315 & 5.666$\pm$0.069 & {\sc this work}\\
AzTEC & 1100\um\   & mJy & --- & --- & 4.6$\pm$1.2 & \citet{aretxaga11a} \\
MAMBO & 1200\um\  & mJy & --- & --- & 4.9$\pm$0.9 & \citet{bertoldi07a} \\
ALMA-B6 & 1287\um\ & mJy   & 1.39$\pm$0.09  & 0.31$\pm$0.09   & 2.05$\pm$0.11   & {\sc this work}\\
ALMA-B4 & 2038\um\ & \uJy\ & 556$\pm$83     & (15$\pm$83)      & 630$\pm$74      & {\sc this work}\\
ALMA-B3 & 2880\um\ & \uJy\ & 171.6$\pm$7.9   & 24.1$\pm$7.9    & 190.9$\pm$8.5   & {\sc this work}\\
ALMA-B3 & 3287\um\ & \uJy\ & 79.8$\pm$5.6    & 9.0$\pm$6.4     & 103.5$\pm$7.5   & {\sc this work}\\
VLA-3\,GHz & 10\,cm & \uJy\ & 7.34$\pm$2.29 & (3.69$\pm$2.26)          & 10.6$\pm$4.1 & \citet{smolcic17a} \\
\hline\hline
\end{tabular}
\label{tab:photometry}
\vspace{1mm}
\begin{minipage}{\textwidth}
{\small {\bf Notes.} Measurements with $<$3$\sigma$ significance are
  enclosed in parentheses, denoting a formal non-detection; this
  includes measurements that have negative flux density consistent
  with no detection. Note that optical/near-infrared
    constraints for each component A and B are measured using a
    0\farcs6 aperture centered on the ALMA 870\um\ resolved
    components.
}
\end{minipage}
\vspace{2mm}
\end{table*}

\section{Data \&\ Observations} \label{sec:observations}

\subsection{Source Selection \&\ Prior Identification}

The galaxy, \sourcelong, first appeared in the literature in
\citet{bertoldi07a} as ``ID9'' detected by the Max-Planck Millimeter
BOlometer (MAMBO) instrument at the IRAM 30\,m telescope at a
wavelength of 1.2\,mm with S$_{\rm
    1.2}$=4.9$\pm$0.9\,mJy.  We adopt the shorthand name
\source\ throughout this paper.  Plateau de Bure Interferometer
imaging of \source\ exists at 1\,mm with 4\,$\sigma$ significance (its
analysis was included in the PhD thesis of Manuel Aravena,
2009\footnote{Aravena (2009) is available at:
  http://hss.ulb.uni-bonn.de/2009/1687/1687.pdf}).  The redshift was
not known at the time.  The detection was independently corroborated
by \citet{aretxaga11a} as ``AzTEC/C148'' using the AzTEC instrument on
the the Atacama Submillimeter Telescope Experiment (ASTE) at 1.1\,mm
with S$_{\rm 1.1}$=4.6$\pm$1.2\,mJy, in agreement with the
  earlier MAMBO measurement.
The source was then later identified in the SCUBA-2 850\um\ map of
\citet{casey13a} as ``850.43'' (with  S$_{\rm
    850}$=5.55$\pm$1.11\,mJy and no corresponding
450\um\ counterpart) and further as ``COS.0059'' in \citet{geach17b}
 with S$_{\rm 850}$=5.84$\pm$0.87\,mJy.

\source\ has no clear counterpart from either {\it Spitzer} or {\it
  Herschel} in the range
24\um--500\um\ \citep{le-floch09a,lutz11a,oliver12a}.  The lack of
detection in these bands implies that the SED traces unusually cold
dust or, alternatively, a very high redshift solution.  This prompted
a number of teams to pursue ALMA follow-up observations of the source,
including the 3\,mm spectral scan presented by \citet{jin19a}.

Our interest in \source\ stems from the new ALMA 2\,mm blank-field in
the COSMOS field (Cycle 6 program 2018.1.00231.S, PI Casey).  The
scientific objective of the 2\,mm blank-field map is to constrain the
volume density of DSFGs at $z\simgt4$.  This is made possible because
2\,mm detection is an effective way to `filter out' lower redshift
DSFGs at $1\simlt z\simlt3$ as detailed in the modeling work of
\citet{casey18a,casey18b}.  Blank-field maps at shorter wavelengths
(e.g. 870\um\ and 1.1\,mm) identify more sources than at 2\,mm per
given solid angle, but such work then suffers from the ``needle in the
haystack'' problem of identifying which sources sit at $z\simgt4$
\citep[e.g. as described in][]{casey19a}.  Analysis of the 2\,mm
blank-field map dataset will follow in a later paper.

\source\ was the brightest source identified in the first
9.4\,arcmin$^{2}$ of delivered 2\,mm map data, and the ratio of
850\um\ flux density to 2\,mm flux density ($S_{\rm 850\mu\!m}/S_{\rm
  2mm}=8.3\pm0.9$) implied a high redshift solution, where a higher
value ($S_{\rm 850\mu\!m}/S_{2mm}=15\pm3$) would be expected for DSFGs
at $z\sim1-4$.  An independent analysis of data from Cycle 5 program
2017.1.00373.S (PI Jin) identified a 4$\sigma$ emission line
consistent with the measured redshift solution as well as other
possible high-$z$ solutions (and corresponding candidate
$\sim$4$\sigma$ emission peaks), which led to the proposal for the
data described herein.  This line and spectroscopic redshift have
since been reported in \citet{jin19a} based on the identification of
the tentative 101\,GHz line as \cof\ and a line at the edge of ALMA
band 3, $\sim$84\,GHz as \coe.  Our independent analysis of the same
data did not lead to a significant detection of the \coe\ line.
Because the 84\,GHz line sits at the very edge of ALMA band 3 (whose
lower limit frequency is 84\,GHz) and at low S/N, additional tunings
were needed to elucidate the redshift solution and characterize
\source.

\begin{figure}
\centering
\includegraphics[width=0.99\columnwidth]{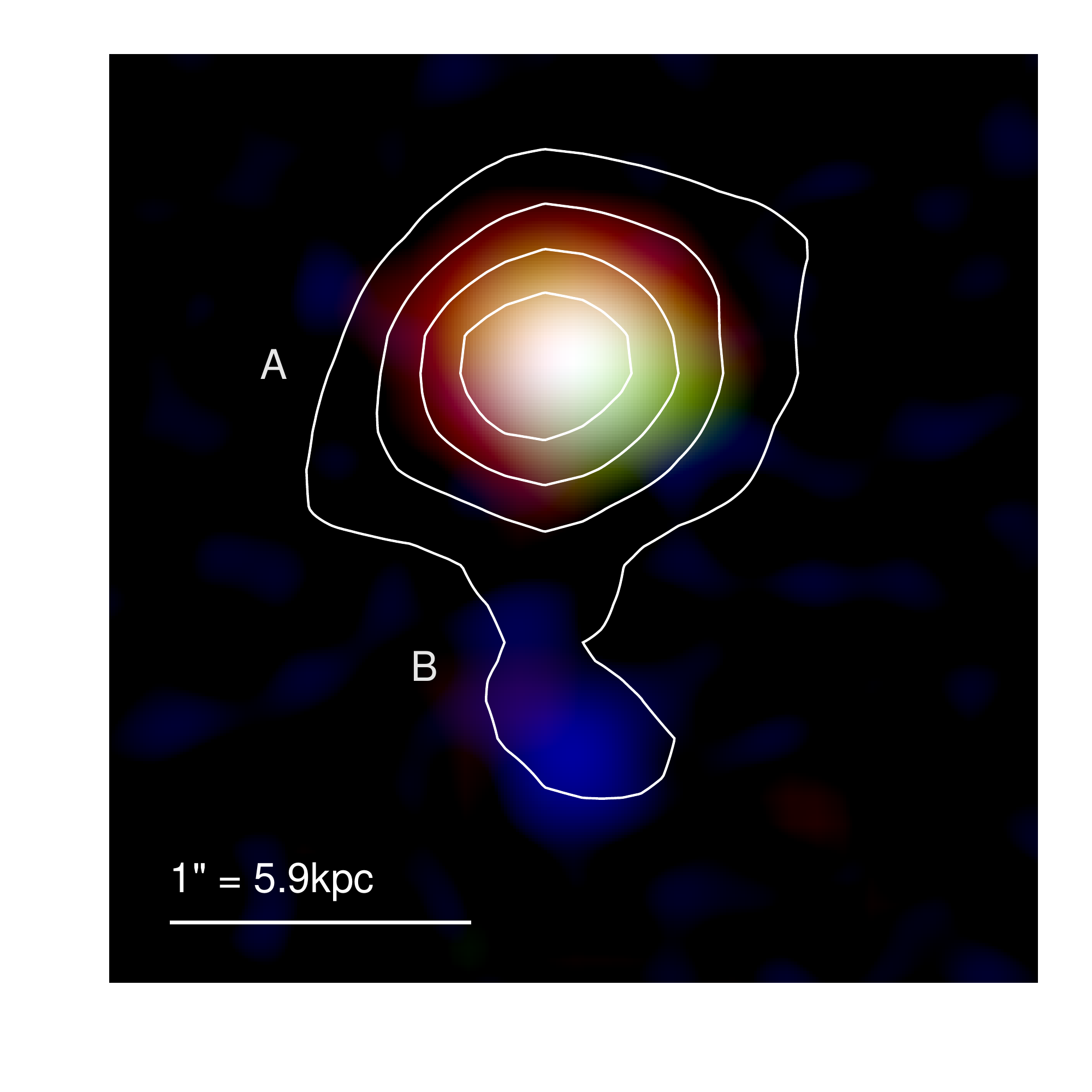}
\caption{A three-color rendition of the dust continuum emission in
  \source: blue represents 870\um\ emission, green is 1.3\,mm, and red
  is 3\,mm emission with similar beamsizes.  Briggs
    weighting with robust=0 is used in all bands with beamsizes of
    0\farcs36, 0\farcs75, and 0\farcs65, respectively.  Integer multiples of
  $\sigma$ above three are shown in contours for the \cof\ line
  emission (at intermediate spatial resolution using Briggs weighting)
  in context.  The northern source is component A and southern
  component B of \source. The \cof\ emission in component B is
  spatially coincident with the 870\um\ dust emission.  The difference
  in millimeter color between the two components is real; in other
  words, component B would have been detected at 3\,mm in dust
  continuum if it had a similar SED to component A.}
\label{fig:tricolor}
\end{figure}

\subsection{ALMA Data}

Observations with the Atacama Large Millimeter/submillimeter Array
(ALMA) were obtained under program 2018.1.00037.A split into three
scheduling blocks tuned to three different frequencies: two in band 3
(3\,mm) and one in band 7 (870\um). The frequencies were chosen to
secure the redshift of \source\ which was determined to have multiple
viable solutions\footnote{These observations were planned before the
  results of \citet{jin19a} were known, though we did consider
  $z=5.85$ as one of four possible redshift solutions.}  at $4\simlt
z\simlt9$. The ALMA data were reduced and imaged using the Common
Astronomy Software Application\footnote{http://casa.nrao.edu} ({\sc
  CASA}) version 5.4.0 following the standard reduction pipeline
scripts and using manually defined clean boxes during the cleaning
process.  For band 7 observations, \source\ is detected at very high
signal-to-noise (111$\sigma$) such that we performed self-calibration.
Also in band 7, a few noisy channels in one spectral window were
identified and flagged in the frequency range 331.35--333.33\,GHz
after visual inspection of the calibrated data. No additional flagging
was required for band 7 or band 3 observations.

Band 7 observations covered frequencies 329.5--333.5\,GHz and
341.5--345.5\,GHz.  They were taken on 2019-05-02 in the C43-4
configuration, with a synthesized beam of $0\farcs36\times0\farcs30$
(using natural weighting), a total integration time of 6383\,s, and a
mean precipitable water vapor (PWV) of 0.9\,mm.  The continuum RMS
reached over the 7.5\,GHz bandwidth is 26.9\,\uJy/beam.  We explored
different visibility weights for imaging, using both natural and Briggs
weighting (robust=0.0--0.5), the latter to spatially resolve the
distribution of dust in the primary components of \source.

\begin{figure*}
\centering
\includegraphics[width=1.99\columnwidth]{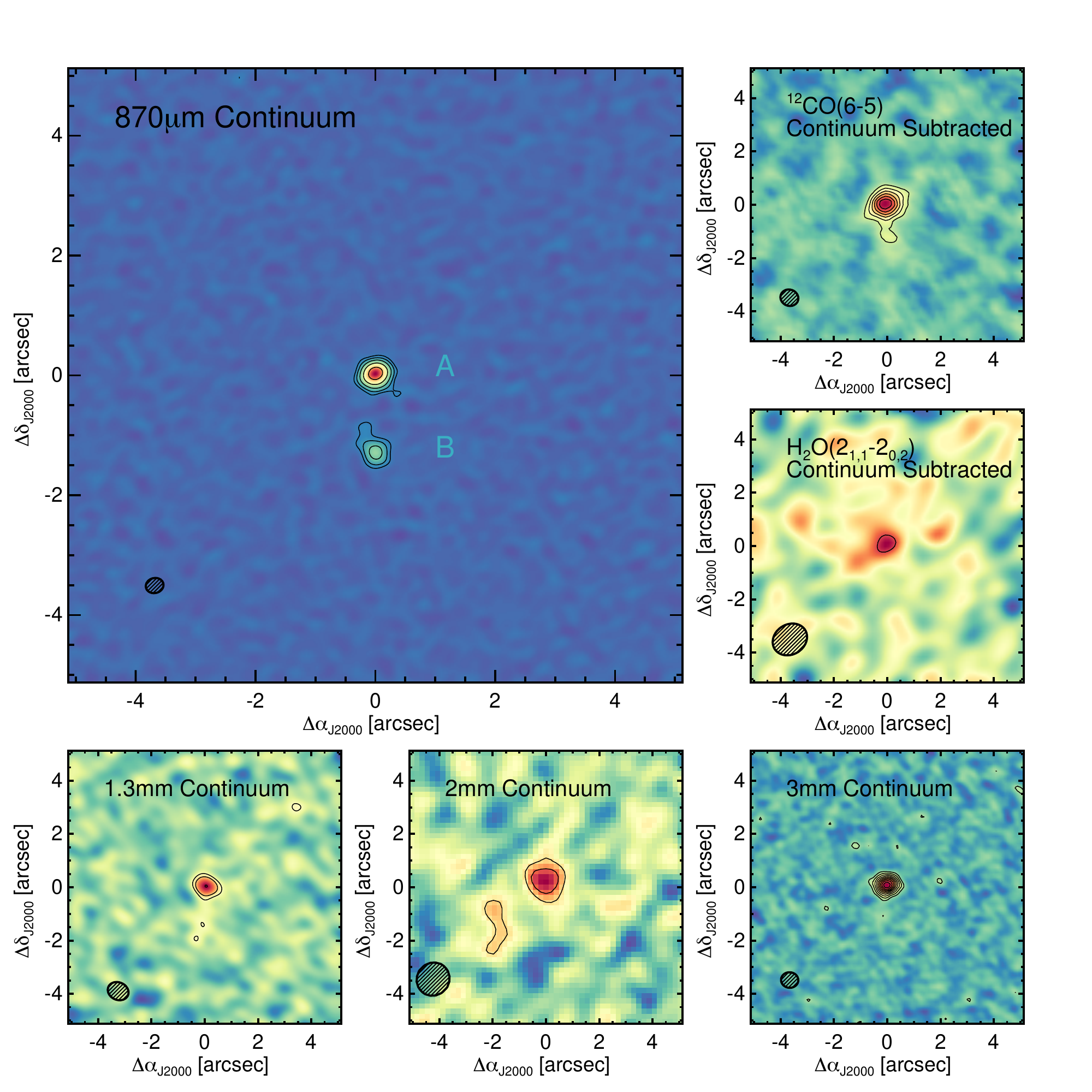}
\caption{Ten arcsecond image cutouts of \source\ from ALMA datasets,
  including 870\um, 1.3\,mm, 2\,mm, 3\,mm continuum (bands 7, 6, 4,
  and 3), as well as continuum-subtracted moment-0 maps of the
  \cof\ and \htwoo\ lines. At 870\um, contours follow five times
  integer powers of two (from 5--80$\sigma$); all other maps follow
  odd integer multiples of $\sigma$ between 3--21$\sigma$.  The peak
  signal-to-noise is 111$\sigma$ at 870\um\ continuum, 20.7$\sigma$ at
  3\,mm continuum, 6.7$\sigma$ at 2\,mm, 9.2$\sigma$ at 1.3\,mm,
  14.7$\sigma$ in the \cof\ moment-0 map and 4.3$\sigma$ in the
  \htwoo\ moment-0 map.  In reference to the 870\um\ image, the
  northern, brighter source is component A and the southern, fainter
  source is component B.  While component A is significantly detected
  in all maps, component B is only significantly detected
  ($>$3$\sigma$) at 870\um\ and in \cof\ (though there are marginal
  detections at 1.3\,mm and 3\,mm).}
\label{fig:maps}
\end{figure*}

Band 3 data were obtained in two tunings.  The first covers
frequencies 86.6--90.3\,GHz and 98.6--102.4\,GHz.  These data were
taken on 2019-04-30 and 2019-05-01 in C43-4 with a total integration
time of 14668\,s, resolution of $1\farcs19\times1\farcs09$ (using
natural weighting), and an average PWV ranging from 1.3--2.3\,mm.  The
RMS reached over a 50\,\kms\ channel width was 0.124\,mJy/beam.  The
second band 3 tuning covers frequencies 94.8--98.5\,GHz and
106.6--110.4\,GHz.  These data were taken on 2019-04-30, 2019-05-01
and 2019-05-02, a spatial resolution of of $1\farcs14\times0\farcs93$
(using natural weighting), a total integration time of 15010\,s, and a
mean PWV ranging from 0.9--2.0\,mm.  The RMS reached for a
50\,\kms\ channel width was 0.088\,mJy/beam.  All band 3 data
presented in this paper also are co-added in the visibility plane with
the archival spectral scan from \citet{jin19a} (2017.1.00373.S) which
contributes a total of $\sim$1500\,s integration time to the total
(9\%).  Our 3\,mm continuum flux density is consistent
  with the measurement from the \citeauthor{jin19a} data. In an
effort to spatially resolve the components of \source, we explore
different weightings with different synthesized beamsizes.  The
overall continuum RMS achieved in the co-addition of all the band 3
data is 5.2\,\uJy\ using natural weighting to maximize the line
signal-to-noise ratio.

Band 6 continuum data also exists for \source\ from the 2016.1.00279.S
program (PI: Oteo) which achieved a continuum sensitivity of
0.16\,mJy/beam at a representative frequency of 233\,GHz; the
synthesized beamsize in the band 6 data is 0\farcs81$\times$0\farcs68
(with Briggs weighting and 
robust=0.0).  Band 4 continuum data, from our separate
2\,mm blank-field map program, 2018.1.00231.S (PI: Casey) has a
continuum sensitivity of 0.11\,mJy/beam, representative frequency of
147\,GHz, and synthesized beam of 1\farcs83$\times$1\farcs43 (natural
weighting).  More analysis of the band 4 data will be presented in a
forthcoming work.  We use Briggs weighting with
  robust=0.0 where the signal-to-noise is sufficiently high to provide
  improved spatial resolution, while we use natural weighting to
  measure sources' integrated flux densities, especially when the
  detection signal-to-noise ratio is near the 5$\sigma$ threshold.

 Analysis of the band 7 data reveals two distinct point
  sources separated by $\sim$1$''$ oriented in a North-South direction
  as shown in Figure~\ref{fig:tricolor} and Figure~\ref{fig:maps}. We
  call the northern, brighter source component A (or \source--A) and
  the southern fainter source component B (or \source--B).

\subsection{Ancillary Archival COSMOS Datasets}

\source\ sits in the central portion of the Cosmic Evolution Survey
Field (COSMOS) covered by the Cosmic Assembly Near-infrared Deep
Extragalactic Legacy Survey \citep[CANDELS;][]{koekemoer11a,grogin11a},
and thus benefits from some of the deepest ancillary data available.
This source has no counterpart in the deep imaging catalog of
\citet{laigle16a}.  The source is detected in the deep S-CANDELS {\it
  Spitzer} IRAC data \citep{ashby15a} and appears to be marginally
resolved in the north-south direction, consistent with the positions
and orientation of components A and B with respect to one another.
There is a marginal detection ($\sim$3$\sigma$) of a portion of the
source in the deep {\it HST} F160W imaging data near component
A. However, using a 0$\farcs$6 extraction aperture centered on the
ALMA 870\um\ dust map reduces this potential marginal detection to
$<$1$\sigma$ significance.  There is also a detection of a faint
source 1$''$ to the south of component B in UltraVISTA $Ks$-band, {\it
  Hubble} F125W and F160W imaging, though we believe it is
unassociated with \source\ based on the different optical/infrared
colors (e.g. Ks-band magnitude of $K_{s}=26.36\pm0.35$
  yet no associated IRAC emission) and lack of ALMA counterpart in
the extraordinarily deep 870\um\ image. \source\ is not detected in
the {\it Spitzer} 24\um\ imaging \citep{le-floch05a}, nor {\it
  Herschel} PACS 100\um, 160\um\ \citep{elbaz11a}, or SPIRE 250\um,
350\um\ or 500\um\ \citep{oliver12a}, which would be expected for
sources of similar 850\um\ flux densities at $z\simlt2-3$
\citep{casey12a,casey12b,gruppioni13a}.  Note that \citet{jin19a}
report photometry for this source in the {\it Herschel} SPIRE bands
using the ``super-deblended'' extraction technique \citep{jin18a},
although examination of the {\it Herschel} map shows no detection or
contamination by nearby neighbors; we adopt upper limits for the SPIRE
bands instead. Our upper limits come from the confusion noise RMS
which dominates the uncertainty of flux calibration at low S/N
\citep[][upper limits in the {\it Herschel} PACS bands are limited by
  instrumental noise, \citealt{lutz11a}]{nguyen10a}.  \source\ is not
detected in the deep 1.4\,GHz radio imaging of \citet{schinnerer07a},
and though not formally detected above the 5\,$\sigma$ threshold in
the 3\,GHz VLA map, there is a 3.2$\sigma$-significance peak near
\source-A at 3\,GHz \citep{smolcic17a}.  There is no X-ray
detection. Across all measured datasets, \source\ is only detected
above $>3\sigma$ significance in seven of 30$+$ different wavebands.
The constraining photometric data are presented in
Table~\ref{tab:photometry}.

We conclude that \source\ is unlikely to be strongly gravitationally lensed.
This is due to the lack of foreground galaxies detected at other
wavelengths in the optical and near-infrared.  The nearest possible
lensing galaxy is offset 4\farcs1 to the north, has a photometric
redshift of $z=2.4$ and estimated stellar mass of
4$\times$10$^{9}$\,\msun; assuming a halo mass of
4$\times10^{11}$\,\msun, this would then lead to a maximum value of
lensing magnification of $\mu=1.15-1.3$ based on conservative
assumptions as to the lensing Einstein radius.  Strong gravitational
lensing by foreground galaxies does affect several other well-known
high-$z$ DSFGs, including the three DSFGs known to sit at higher
redshifts than \source: G09\,83808 at $z=6.03$ \citep{zavala18a},
HFLS3 at $z=6.34$ \citep{cooray14a}, and SPT0311 at $z=6.9$
\citep{strandet17a}.  All of these systems have bright
optical/near-infrared sources within 2$''$ of the millimeter source
center.  This leads us to conclude that \source\ is the
highest-redshift unlensed DSFG identified to-date.

\section{Results}\label{sec:results}

Joint analysis of our ALMA data leads to a spectroscopic confirmation
of \source\ at $z=5.850$ through the detection of \cof\ at
14.7$\sigma$ significance and \htwoo\ at 4.3$\sigma$ significance.
The CO line is detected in
  both components A and B, while the \htwoo\ is only detected in
  component A.  This implies component B only has a single spectral
  line detection.  However, we determine that it's extremely unlikely
  for component B to sit at a redshift that is physically unassociated
  with component A due to their proximity on the sky, similarity in
  optical/near-infrared photometry, and detection of \cof\ emission.
The aggregate
photometry for both components is  given in
Table~\ref{tab:photometry}.

Figures~\ref{fig:tricolor} and \ref{fig:maps} shows the ALMA continuum
and line moment-0 maps of \source\ overlaid together and individually.
The 870\um\ and 3\,mm maps use Briggs weighting (robust\,=\,0) to
maximize spatial resolution; 1.3\,mm, \cof\ moment-0 and 2\,mm
continuum maps are shown with a weighting between Briggs and natural
(robust\,=\,0.5) and the \htwoo\ map is shown using natural weighting
(robust\,=\,2) to maximize line signal-to-noise.
Figure~\ref{fig:spec} shows the aggregate ALMA band 3 spectrum of
\source\ in the range 84--110\,GHz.  What follows is analysis of the
detection of the spectral features \cof\ and \htwoo, a discussion of
continuum-derived properties, then physical characteristics derived
from SED fitting.

\subsection{Millimeter Spectral Line Measurements}\label{sec:3mmresults}

\begin{figure*}
\centering
\includegraphics[width=1.99\columnwidth]{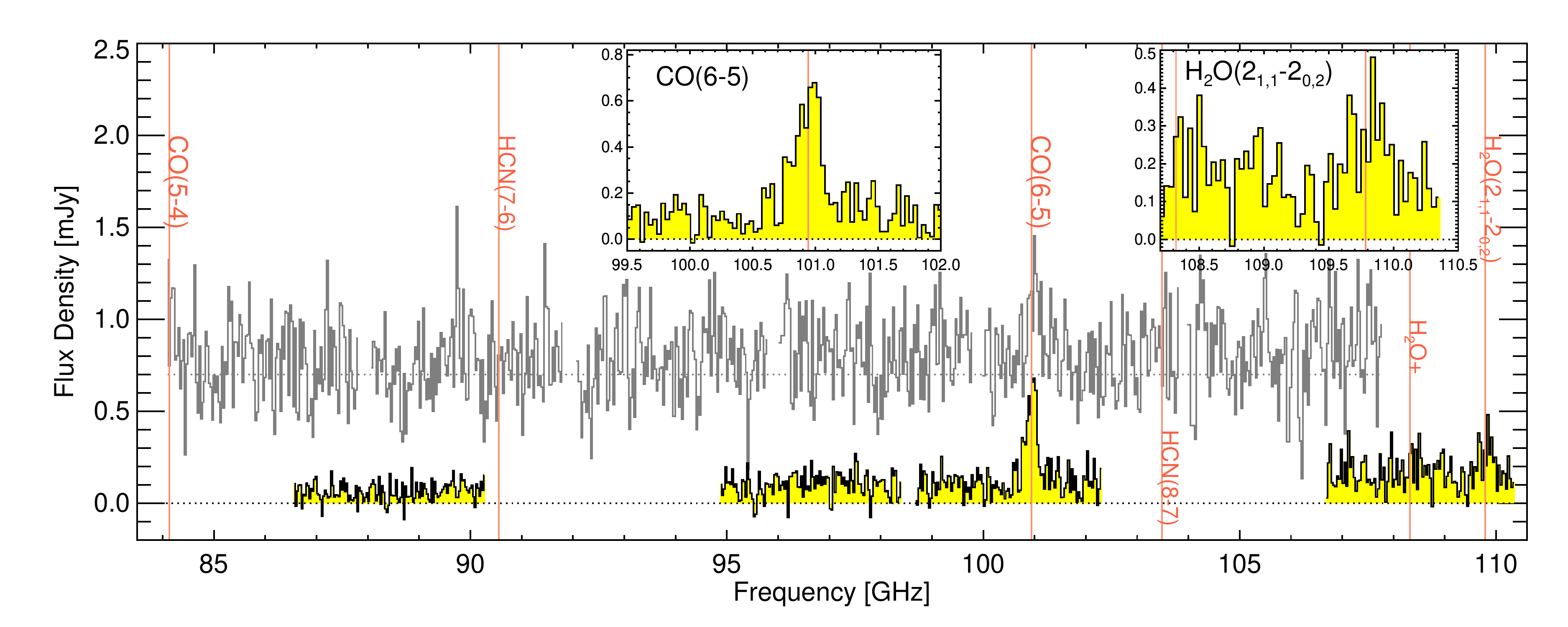}
\caption{The aggregate band 3 spectrum of \source\ from 84--110\,GHz
  extracted over both components A and B.  The original spectrum of
  \citet{jin19a} is shown in grey, offset by 0.7\,mJy, with the
  identification of \coe\ at 84.2\,GHz and \cof\ at 101\,GHz.  Our
  data is shown in yellow histogram and have confirmed the detection
  of the \cof\ line at 101\,GHz and detection of the \htwoo\ line at
  109.8\,GHz, confirming the redshift as $z=5.850$.  Vertical red
  lines mark the expected frequencies of CO and dense gas tracers in
  the observed frequency range.  Inset plots zoom in on the line
  detections at 101\,GHz and 109.8\,GHz.}
\label{fig:spec}
\end{figure*}

Figure~\ref{fig:spec} shows the full band 3 dataset for \source\ in
context; the gray background spectrum is the spectral scan from
\citet{jin19a}, with reported detection of lines at 101\,GHz and
84.2\,GHz corresponding to \coe\ and \cof.  Our
  independent analysis of the \citeauthor{jin19a} dataset, before
  publication of their reported lines, did not lead to significant
  detection of the 84.2\,GHz line.  Thus, 
our band 3 data (shown in
yellow) were tuned to frequencies that would rule in or out other
possible solutions between $4<z<9$.  The detection of emission
features at 101\,GHz and 109.8\,GHz independently corroborate the
reported redshift solution in \citet{jin19a}.
 Note that our derived redshift for component A
  ($z=5.850$) and B ($z=5.852$) differ slightly from the reported
  redshift in \citeauthor{jin19a} of $z=5.847$.  We attribute this to
  the difference in signal-to-noise on the \cof\ feature.

The aggregate band 3 continuum has
  a flux density of 131.4$\pm$5.9\,\uJy\ [20.7$\sigma$ significance]
in the full bandwidth and in many individual channels of our dataset,
thus analysis of molecular line emission requires continuum-subtracted
data.  Note that in Table~\ref{tab:photometry}, the band
  3 continuum flux density is split into two independent measurements
  by frequency, given the high signal-to-noise on the aggregate
  dataset.

The integrated 
\cof\ 
 line flux is 
  0.48$\pm$0.03\,Jy\,\kms\ [14.7$\sigma$ significance]
and the \htwoo\ line flux is
  0.09$\pm$0.02\,Jy\,\kms\ [4.3$\sigma$  significance].  Components A and B are only
distinguishable in band 3 data when using Briggs weighting
(robust=0.0), as natural weighting results in a synthesized beam
slightly larger than the 1$''$ separation between the sources; the
disadvantage of Briggs weighting is the potential for resolving out
emission.  Using Briggs weighting, Component A 
 has an integrated \cof\ line flux of
  0.43$\pm$0.03\,Jy\,\kms\ [13.3$\sigma$ significance] 
while Component B has a line flux of
  0.07$\pm$0.02\,Jy\,\kms\ [3.8$\sigma$
  significance].  The \cof\ and 870\,\um\ continuum are spatially
aligned for Component B (see Figure~\ref{fig:tricolor}).  The
\htwoo\ detection only corresponds to component A.
  Within measurement uncertainties, we find that the sum of band 3
  measurements of component A and component B separately (using Briggs
  weighting and robust=0.0) are in agreement with the total integrated
  quantities as measured with natural weighting.

\begin{figure}
\centering
\includegraphics[height=0.99\columnwidth]{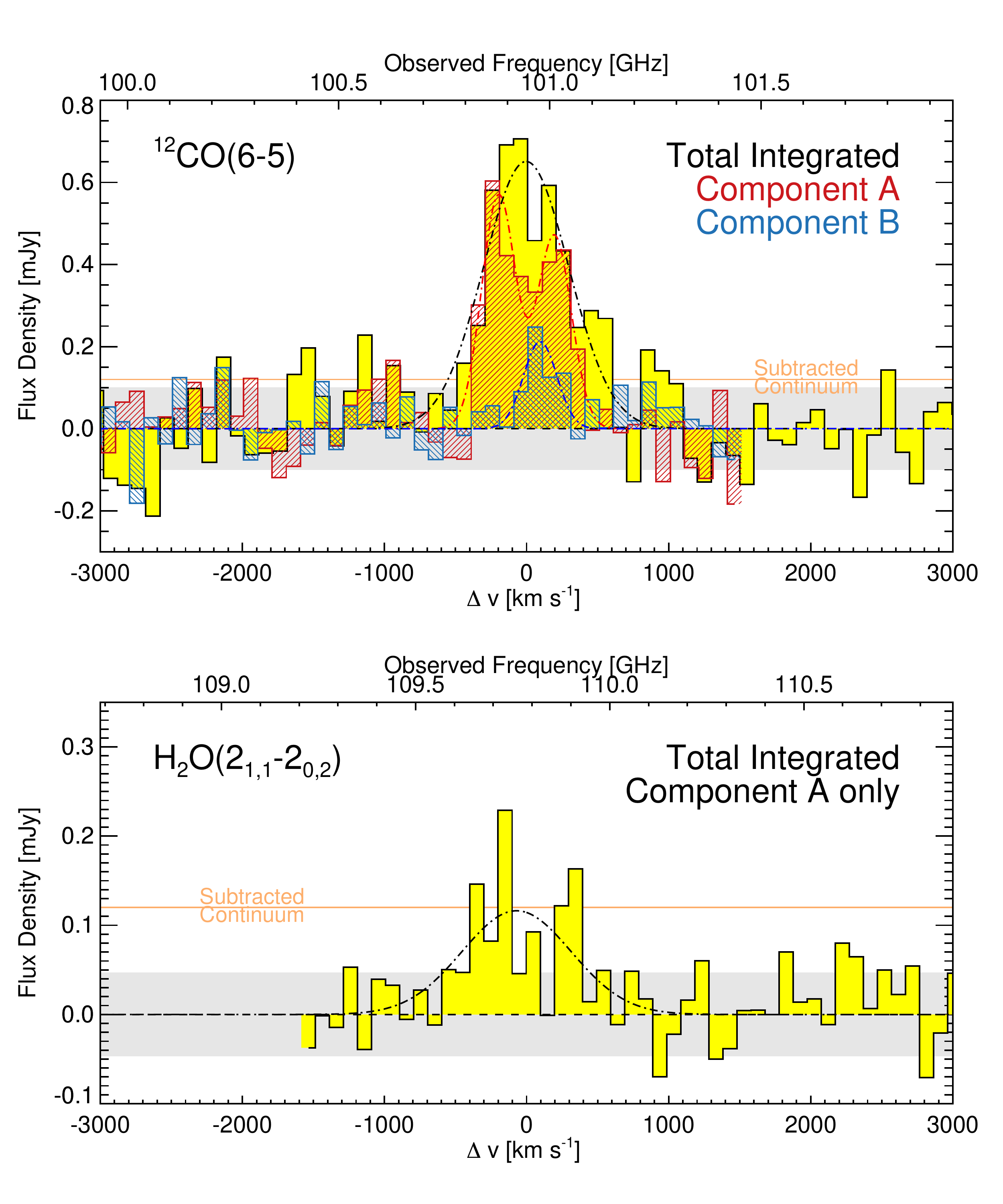}
\caption{Continuum-subtracted source spectra of \source\ in the
  \cof\ and \htwoo\ lines with velocity relative to central redshift
  of $z=5.850$.  The integrated spectrum (filled yellow histogram) is
  analogous to that shown in Figure~\ref{fig:spec}, 
    i.e. spectrum from naturally weighted band 3 data, but with
  continuum emission subtracted (the orange line indicates the level
  of the continuum).  The RMS per channel is indicated by the gray
  horizontal stripe.  The \cof\ line is then further separated into
  two components A and B using the highest spatial-resolution
  reduction (as shown in Figure~\ref{fig:maps}  using
    Briggs-weighted data with robust=0.0).  The coaddition of the
  spectra of component A and B are, within uncertainty, in agreement
  with the total integrated spectrum from the improved-sensitivity
  weighting, suggesting very little \cof\ emission is resolved out.
  The integrated line signal-to-noise ratio for \cof\ is 13.3$\sigma$
  in component A, and 3.8$\sigma$ in component B.  The \htwoo\ line is
  detected at 4.3$\sigma$ in component A only (shown
    here is data with natural weighting where the two components are
    not spatially resolved).}
\label{fig:specdetail}
\end{figure}

Figure~\ref{fig:specdetail} shows the line spectra for both \cof\ and
\htwoo, with \cof\ broken down into the two components.  While the
total line emission appears roughly Gaussian, the spectrum of
component A alone appears double peaked and possibly indicative of
rotation. This suggestive rotation is also seen in the
position-velocity diagram shown in Figure~\ref{fig:kinematics}, as the
source is resolved across $\sim$2.4 beams.

The integrated line fluxes are given in Table~\ref{tab:physical}.  We
measure the FWHM and estimate uncertainties of both the \cof\ and
\htwoo\ features by using Monte Carlo simulations where noise is
injected and the line-width remeasured.  For single-profile Gaussian
fits to the integrated line luminosities, we measure widths of
700$\pm$70\,\kms\ in \cof\ and 900$\pm$200\,\kms\ in \htwoo.  When
analyzing the data for components A and B separately, we measure
single-profile Gaussian widths of 260$\pm$40\,\kms\ for component A
and 280$\pm$130\,\kms\ for component B in \cof.  However, we note that
component A is best fit by a double Gaussian separated by
400\,\kms\ and individual line widths FWHM=370\,\kms.

\subsection{\htwoo\ Emission}

The \htwoo\ line at rest-frame 752\,GHz is a medium-excitation
($E_{\rm up}\,=\,136$\,K) transition of para-H$_{2}$O, commonly seen
in emission in galaxies as a tracer of dense
($n(H)\sim10^{5}-10^{6}\,cm^{-3}$) star-forming gas in the ISM
\citep{gonzalez-alfonso10a,liu17b,jarugula19a,apostolovski19a,yang19a}.
Though rare in the gas phase of non-star-forming molecular clouds
\citep{caselli10a}, water is the third-most common molecule after
H$_{2}$ and CO in shock-heated regions of the ISM that trace
star-forming regions \citep{bergin03a}.  Furthermore, the velocity
structure of H$_{2}$O emission in nearby galaxies tends to mirror that
of CO \citep{liu17b}, suggesting that water is widespread throughout
the bulk molecular gas reservoir of galaxies.  This particular
transition tends to be relatively bright compared to most water
emission features.

The continuum-subtracted line luminosity of the \htwoo\ feature is
(3.6$\pm$0.8)$\times$10$^{7}$\,\lsun, precisely on the \lir--$L_{\rm
  H_{2}O}$ relation found in \citet{yang13b}  using our
  best-constrained \lir\ for component A from \S~\ref{sec:irsed}; this
corroborates earlier results that suggest H$_{2}$O might be a
particularly good star-formation tracer.  Furthermore, the ratio of
line flux between \htwoo\ and \cof\ is $\sim$30\%, consistent to
within 10\%\ of the composite DSFG millimeter spectrum from the SPT
survey \citep{spilker14a}.

\subsection{Dust Mass}

Dust continuum detections on the Rayleigh-Jeans tail of blackbody
emission can be used to directly infer \source's total dust mass (and
also ISM mass by proxy, as discussed in the next subsection). Dust
mass is proportional to dust temperature and flux density along the
Rayleigh-Jeans tail where dust emission is likely to be optically thin
(at $\lambda_{\rm rest}\simgt300$\,\um).  As we discuss
  later in \S~\ref{sec:irsed}, we estimate that this is a safe
  assumption to make in the case of \source, where we do not think the
  SED is optically thick beyond $\lambda_{\rm
    rest}\approx300$\,\um\footnote{If the SEDs were optically thick at
    these wavelengths, the dust mass would be underestimated using
    this technique.}.  Because \source\ sits at a relatively high
redshift, cosmic microwave background (CMB) heating of the dust is
non-negligible and the subsequent measurement of dust mass is
impacted.

For the general case of a galaxy at sufficiently high-$z$ like
\source, dust mass can be calculated using the following:
\begin{equation}
\begin{split}
M_{\rm dust} = &
\frac{S_{\rm \nu_{\rm obs}}D_{\rm L}^{2}(1+z)^{-(3+\beta)}}{\kappa(\nu_{\rm ref})B_{\nu}(\nu_{\rm ref},T_{\rm dust})}\bigg(\frac{\nu_{\rm ref}}{\nu_{\rm obs}}\bigg)^{2+\beta}\bigg(\frac{\Gamma_{\rm RJ(ref,0)}}{\Gamma_{\rm RJ}}\bigg) \\
& \times \bigg(1-\frac{B(\nu_{\rm rest},T_{\rm CMB}(z))}{B(\nu_{\rm rest},T_{\rm dust})}\bigg)^{-1}\\
\end{split}
\label{eq:dustmass}
\end{equation}
Here, observations are acquired at $\nu_{\rm obs}$
  (measured in Hz) with flux density $S_{\nu_{\rm obs}}$
 (measured in erg\,s$^{-1}$\,cm$^{-2}$\,Hz$^{-1}$), and
  $\nu_{\rm rest}=\nu_{\rm obs}(1+z)$.  The frequency at which the
dust mass absorption coefficient is known is $\nu_{\rm ref}$; for
example, a value of
$\kappa(450\mu\!m)\,=\,1.3\pm0.2$\,cm$^{2}$\,g$^{-1}$
\citep{li01a,weingartner01a}, which is observed-frame 3\,mm at $z=6$.
$D_{\rm L}$ is the luminosity distance (converted to
  cm) and $B_{\nu}$ is the Planck function evaluated at a given
frequency and for a given temperature in units of
  erg\,s$^{-1}$\,cm$^{-2}$\,Hz$^{-1}$.  For example, $B(\nu_{\rm
    rest},T_{\rm CMB}(z))$ is the planck function evaluated at
  $\nu_{\rm rest}$ for the temperature of the CMB at the measured
  redshift $z$. M$_{\rm dust}$ is in units of $g$ which can be
  converted to \msun.  $\beta$ is the emissivity spectral index,
which we set to $\beta=1.95$ (we derive this value from a fit to our
data in \S~\ref{sec:irsed}). $\Gamma_{\rm RJ}$ represents the
Rayleigh-Jeans (RJ) correction factor, or the deviation from the RJ
approximation, following the framework of \citet{scoville16a}:
\begin{equation}
\Gamma_{\rm RJ}(\nu,T_{\rm d},z) = \frac{h\nu(1+z)/kT_{\rm d}}{e^{h\nu(1+z)/kT_{\rm d}}-1}
\end{equation}
and $\Gamma_{\rm RJ(ref,0)}=\Gamma_{\rm RJ}(\nu=\nu_{\rm ref},T_{\rm
  d}=T_{\rm dust},z=0)$.  Here $h$ is the Planck
  constant (6.63$\times$10$^{17}$\,erg\,s) and $k$ is the Boltzmann
  constant (1.38$\times$10$^{-16}$\,erg\,K$^{-1}$). $T_{\rm d}$ is
the galaxy's {\it mass-weighted} dust temperature, not the same
quantity as fit in \S~\ref{sec:irsed}, which is the {\it
  luminosity-weighted} dust temperature.  We adopt a mass-weighted
dust temperature of 25\,K throughout to be consistent with
\citet{scoville16a}.
The last multiplicative factor in Eq.~\ref{eq:dustmass} represents the
correction for suppressed flux density against the CMB background, as
described in \citet{da-cunha13a}.  This factor is a function of
$\nu_{\rm rest}=\nu_{\rm obs}(1+z)$, the CMB temperature at the given
redshift, $T_{\rm CMB}=2.73\,K\,(1+z)$, and the CMB-corrected
mass-weighted dust temperature, as given in Equation~12 of
\citet{da-cunha13a}.
An assumption of this formulation is that the dust (at temperatures
similar to the CMB temperature) is optically thin, which holds in
almost all environments, with exception of the densest cores of local
ULIRGs.

We derive dust masses from our 3\,mm continuum photometry centered on
a wavelength of 3085\,\um\ (rest-frame 450\um).  
We infer a dust mass of (1.3$\pm$0.3)$\times10^{9}$\,\msun\ for
component A and (1.9$^{+1.3}_{-0.8}$)$\times10^{8}$\,\msun\ for
component B.  Note that the sum of these values is $\sim$5$\times$
higher than the dust mass derived for this system in \citet{jin19a},
with the difference attributed to the differences in SED fitting
including best-fit $\beta$ (our dust mass uncertainties
  do not account for the measurement uncertainties in $\beta$).

\subsection{Gas Mass}

We 
derive the mass of molecular gas in each galaxy from
  dust continuum as well as from \cof\ line luminosity.
In both cases, we adopt a value for the CO-to-H$_{2}$
  conversion factor of $\alpha_{\rm
    CO}=6.5$\,M$_\odot$\,(K\,km\,$s^{-1}$\,pc$^2$)$^{-1}$ as in
  \citet{scoville16a}.  This is in-line with the Galactic value and
  accounting for the mass of both H$_{\rm 2}$ and He
  gas\footnote{Without including He, the Galactic value is
    $\sim$4.5\,M$_\odot$\,(K\,km\,s$^{-1}$\,pc$^{2}$)$^{-1}$.}.

We follow the methodology described in the Appendix of
\citet{scoville16a} to derive a gas mass from dust
  continuum, modified to account for CMB heating similar to the
impact on the dust mass calculation:
\begin{equation}
\begin{split}
M_{\rm gas}=&\frac{4\pi D_{\rm
    L}^{2}S_{\nu,obs}}{\alpha(\nu_{\rm 850}) (1+z)^{3+\beta}}
\bigg(\frac{\nu_{\rm 850}}{\nu_{\rm obs}}\bigg)^{2+\beta}
\bigg(\frac{\Gamma_{\rm RJ(ref,0)}}{\Gamma_{\rm RJ}}\bigg)\\ & \times
\bigg(1-\frac{B(\nu_{\rm rest},T_{\rm CMB}(z))}{B(\nu_{\rm
    rest},T_{\rm dust})}\bigg)^{-1}\\
\end{split}
\label{eq:gasmass}
\end{equation}
Here $\alpha_{\rm
  850}=(6.7\pm1.7)\times10^{19}$\,erg\,s$^{-1}$\,Hz$^{-1}$\,\msun$^{-1}$
is the empirically calibrated conversion factor from
850\um\ luminosity to ISM mass from \citet{scoville16a}, which
bypasses use of both the uncertain dust-to-gas ratio and dust mass
absorption coefficient (used to measure dust masses above).  Note that
intrinsic to this calculation is the assumed value of the
CO-to-H$_{\rm 2}$ conversion factor as stated above.  Similar to the
calculation of dust mass, we adopt a single mass-weighted dust
temperature of 25\,K.  $M_{\rm gas}$ as given in
Equation~\ref{eq:gasmass} is in units of \msun.
With this approach we constrain masses of molecular gas using dust
continuum to be $M_{\rm
  gas}$\,=\,(1.4$\pm$0.4)$\times10^{11}$\,\msun\ for component A and
$M_{\rm gas}$\,=\,(1.2$\pm$0.5)$\times10^{10}$\,\msun\ for component
B.

Historically it has been more common to use transitions of CO to infer
the underlying gas mass in galaxies \citep{solomon05a,carilli13a}, yet
it comes with substantial uncertainty, especially when using high-J
transitions like \cof.  High-J transitions of CO tend to trace dense
gas regions of the ISM, which are a relatively poor probe of the
entire molecular gas reservoir of a galaxy.  Nevertheless, here we
offer a calculation of the gas mass from \cof\ as an independent check
against what we have calculated using dust continuum.

We use the \cof\ line luminosity\footnote{This uses the standard
  $\L_{\rm CO}^\prime$ definition as in Equation~3 of
  \citet*{solomon05a}, the first equation of \citet*{carilli13a}, and
  Equation~19 of \citet*{casey14a}.} to derive a molecular gas mass
from \cof.  This requires an assumption as to the value of the gas
excitation spectral line energy distribution (SLED) to convert from
\cof\ to the ground-state \coa\ and then the value of the
CO-to-H$_{2}$ conversion factor \citep{bolatto13a}.  
We assume that \source\ has a CO SLED similar to other high-$z$ DSFGs
in the literature \citep*[summarized in Figure~45 of][including a
  substantial contribution from the compilation of
  \citealt{bothwell13c}; we adopt the blue shaded region from that
  figure as the 1$\sigma$ uncertainty on $I_{\rm CO(6-5)}/I_{\rm
    CO(1-0)}=10^{+30}_{-5}$]{casey14a}.  
 We calculate gas masses of $M_{\rm
  gas}(A)=(3.3\pm1.7)\times10^{11}$\,\msun\ and $M_{\rm
  gas}(B)=(5\pm3)\times10^{10}$\,\msun\ for components A and B
respectively.  These are broadly consistent with, yet more uncertain
than, the dust continuum derived gas masses.

Later in \S~\ref{sec:massbudget} we discuss the implications of the
rarity of this halo on the measured gas mass, in particular the
assumption that $\alpha_{\rm
  CO}=6.5$\,\msun\,(K\,km\,s$^{-1}$\,pc$^{2}$)$^{-1}$. 
  Both calculations of the gas masses account for measurement
  uncertainties in flux density, $\alpha_{\rm 850}$ or $I_{\rm
    CO(6-5)}/I_{\rm CO(1-0)}$, but not uncertainty in $\alpha_{\rm
    CO}$.  If we were to instead take $\alpha_{\rm
  CO}\,=$\,1\,\msun\,(K\,km\,s$^{-1}$\,pc$^{2}$)$^{-1}$, more in line
with measured constraints on low- and high-redshift dusty starbursts
\citep[e.g.][]{downes98a,tacconi08a,bolatto13a}, the gas mass would
scale down proportionally by a factor of 6.5.  This would give us gas
masses of $M_{\rm gas}$\,=\,(2.2$\pm$0.6)$\times10^{10}$\,\msun\ for
component A and $M_{\rm
  gas}$\,=\,(1.8$\pm$0.8)$\times10^{9}$\,\msun\ for component B
 (both scaled down from the dust continuum-estimated gas
  masses).

\subsection{Size Measurements}\label{sec:sizes}

We measure resolved sizes for both components A and B multiple ways to
check consistency.  First we fit Sersic profiles to the two components
of the highest resolution and signal-to-noise data we have: the
self-calibrated 870\um\ data using Briggs weighting with
  robust=0.0. This probes rest-frame $\sim$127\,\um, near the peak of
the long-wavelength SED, and so these sizes trace the star-forming
region in \source\ most closely.  We perform this analysis in the
$uv$-plane following the methodology outlined in \citet{spilker16a}.
Because both components are only marginally resolved, the size is
measured with a fixed Sersic index of $n=1$ consistent with an
exponential disk (though we found that Gaussian sizes, with $n=0.5$,
are consistent with those measured for $n=1$).
We measure circularized half-light radii of $R_{\rm
  e}(A)=$$0\farcs068\pm0\farcs002=408\pm12$\,pc and $R_{\rm e}(B)=0\farcs110$$\pm0\farcs010=660\pm60$\,pc.
 Uncertainties do not account for unconstrained Sersic
  index which was fixed to $n=1$.  We compare these sizes to the
deconvolved two-dimensional Gaussian sizes measured in the image plane
following the methodology of \citet{simpson15a} and \citet{hodge16a};
we find that, though the image plane fits are somewhat sensitive to
image weighting, they are broadly consistent with the $uv$-plane
analysis, with $R_{\rm
  e}(A)=$0\farcs064$\pm$0\farcs004\,=\,380$\pm$30\,pc, and $R_{\rm
  e}(B)=$0\farcs128$\pm$0\farcs021\,=\,760$\pm$130\,pc (both of these
quoted values are for Briggs weighting).
Note that, even though the synthesized beam of these data is larger
than the measured sizes, the very high signal-to-noise enables us to
measure half-light radii sufficiently smaller than the beamsize FWHM.

Though our 3\,mm data are not at the same high signal-to-noise as the
870\um\ data and also have lower spatial resolution, we fit sizes to
both continuum and CO moment-0 maps of component A 
  (using Briggs weighting with robust=0.0) to explore a possible
differences using the different tracers.  Unfortunately, component B
is not detected at sufficiently high signal-to-noise to have
measurable sizes in either 3\,mm continuum or CO.  We measure a 3\,mm
continuum circularized half-light radius in the image
  plane of $R_{\rm
  e}(A)=$0\farcs123$\pm$0\farcs047\,=\,700$\pm$300\,pc, only 1$\sigma$
discrepant with the measured size at 870\um.  The 3\,mm continuum is a
probe of the cold dust in the ISM.  Though we infer that the CO
moment-0 map is marginally resolved, we find that component A is
consistent with both a point source and the measured 3\,mm and
870\um\ sizes; there is significant uncertainty in the CO size due to
the lower signal-to-noise ratio and poor spatial resolution.

We also use the measured sizes to estimate the average dust column
densities within the half-light radius, which further informs the
conditions of the ISM in \source; we adopt the measured
870\um\ $R_{\rm eff}$ sizes due to the high signal-to-noise near the
peak of dust emission and the measured dust masses to calculate
$\Sigma_{\rm M_{\rm dust}}(A)=1400\pm400$\,\msun\,pc$^{-2}$ and
$\Sigma_{\rm M_{\rm dust}}(B)=50\pm30$\,\msun\,pc$^{-2}$ for A and B
respectively.  If a Milky Way type dust is assumed \citep[as tabulated
  in Table~6 of][]{li01a}, these dust mass column densities imply that
the dust SED should likely be opaque to rest-frame wavelengths of
$\sim$200--400\um\ in the case of component A and $\sim$25--70\um\ in
the case of component B.
Similarly, we can estimate the star-formation surface density
 (using the SFRs derived for each component later in
  \S~\ref{sec:irsed}), and arrive at $\Sigma_{\rm
  SFR}(A)=640\pm170$\,\sfr\,kpc$^{-2}$ and
60$\pm$35\,\sfr\,kpc$^{-2}$.  Neither is near the hypothetical
Eddington Limit for starbursts factors of several larger
\citep{scoville01a,scoville03a,murray05a,thompson05a}, though some
have recently pointed out that the limit might be much higher yet
considering starbursts are distributed over some area and are not
point sources.

\begin{figure}
\centering
\includegraphics[width=0.99\columnwidth]{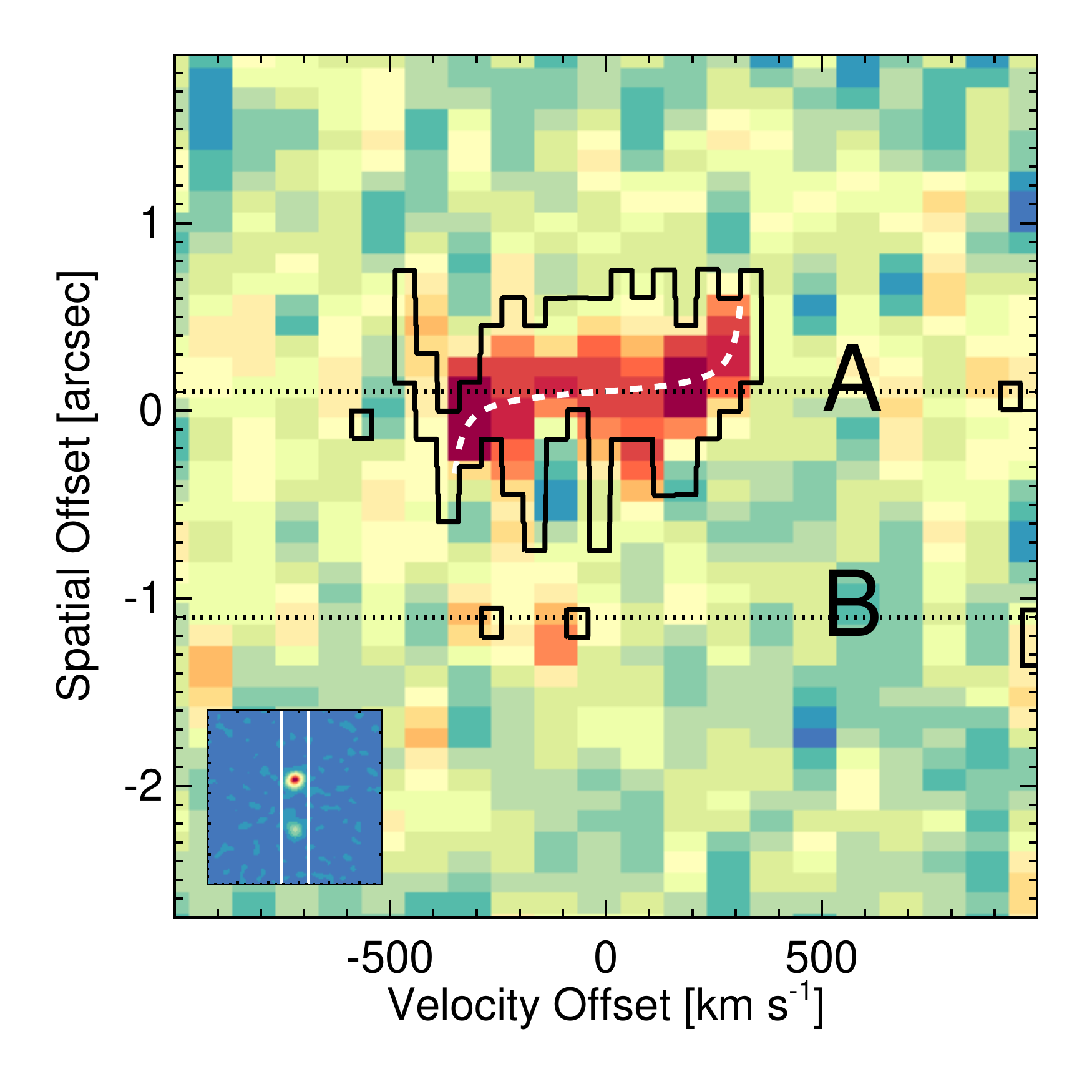}
\caption{A position-velocity diagram of the \cof\ line in the
  highest-resolution Briggs-weighted (robust=0.0) data
  cube overlaid with the 3$\sigma$ significance  black,
    solid contours of a slightly lower resolution processing
  (robust=0.5) of the same data. The
    image color scale is the same as in the on-sky projection in
    Figure~\ref{fig:maps}. Both are extracted using a 0.7$''$-wide
  `slit' with orientation position angle of 0$^o$ as
    shown in the lower left inset plot.  Component A spans a spatial
  extent 0\farcs85$\pm$0\farcs20 (=\,5.0$\pm$1.2\,kpc total extent)
  and velocity $V_{\rm max}\,=$\,350$\pm$50\,\kms.  The
  position-velocity kinematics are suggestive of rotation (white
  dashed line) with $V_{\rm max}=300$\,\kms, though they do not rule
  out more complex interaction dynamics at the given spatial
  resolution.  Component B (spatially offset 1$''$ to the south of
  Component A) is barely detected in this high resolution data cube
  but detected at higher significance at lower resolution and in the
  moment-0 line map.}
\label{fig:kinematics}
\end{figure}

\subsection{Dynamical Mass}

We derive the dynamical masses of \source--A and \source--B using the
\cof\ kinematic profile, comparing a few different methods.  First,
for a galaxy with an unresolved velocity field the dynamical mass is
best estimated by
\begin{equation}
M_{\rm dyn} = \frac{5 R_{\rm e}\sigma^2}{G}
\end{equation}
where $G$ is the gravitational constant, $\sigma$ is the
  measured velocity dispersion of the kinematic feature measured (in
  our case \cof), and $R_{\rm e}$ is the effective circularized
radius and the factor of five is a constant of proportionality
determined to best represent galaxies in the local mass plane
\citep{cappellari06a,toft17a}; this constant does not account for the
inclination angle, $i$.  Correcting for unknown
inclination requires an additional factor of 3/2 (which is the
reciprocal of the expectation value of sin$^{2}i$).

\begin{figure*}
\centering
\includegraphics[width=1.85\columnwidth]{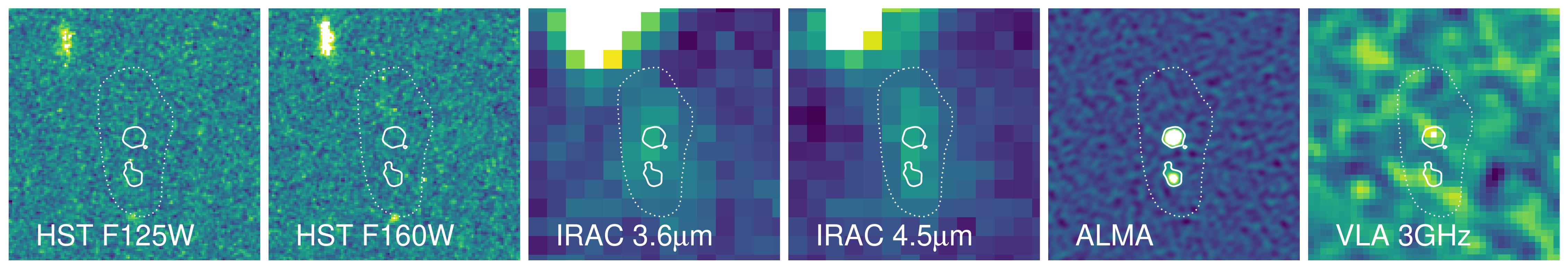}
\includegraphics[width=1.99\columnwidth]{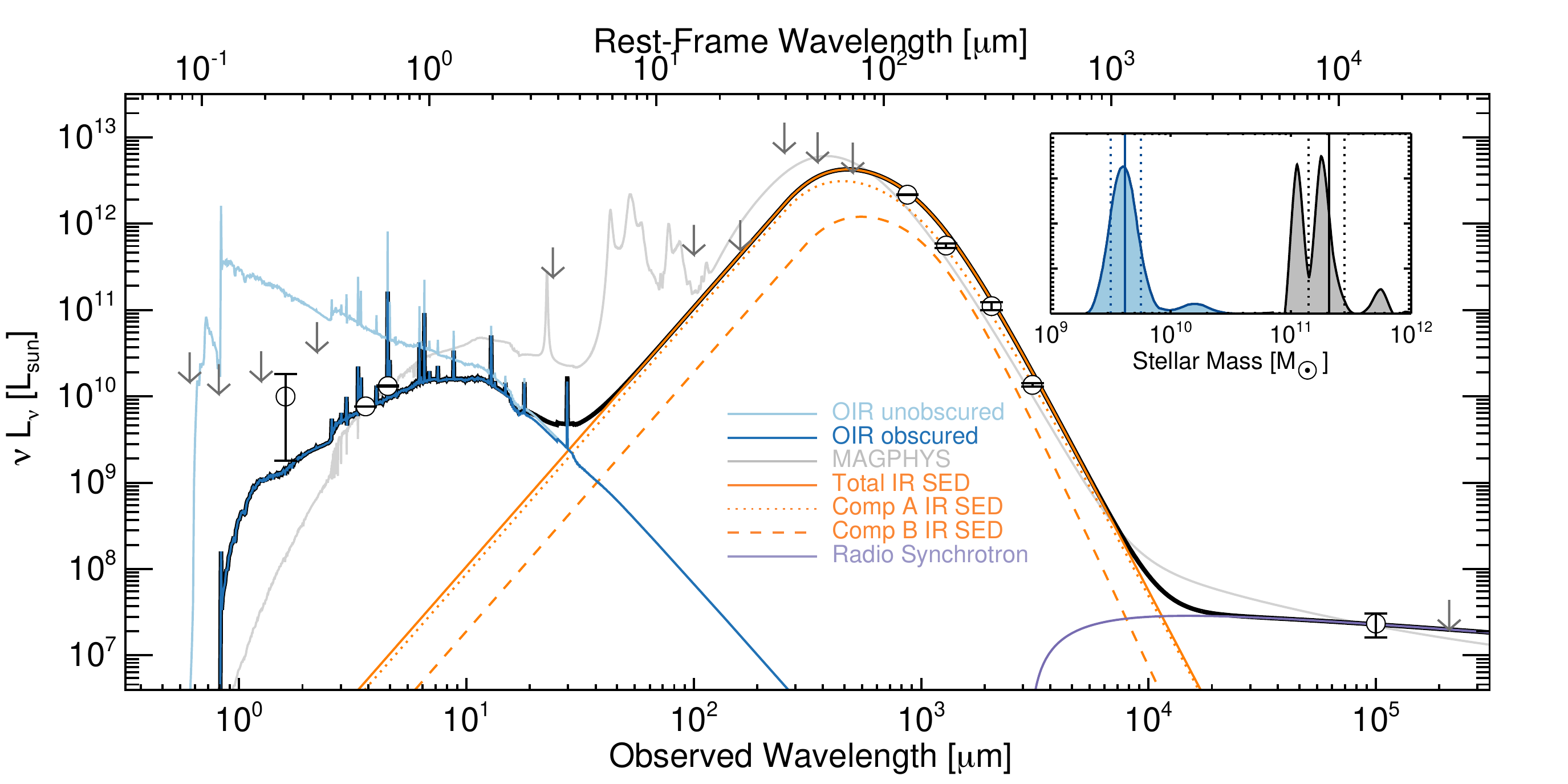}
\caption{At top, 6$''\times$6$''$ cutouts of \source\ in two {\it HST}
  bands, the IRAC bands, ALMA 870\um\ and VLA 3\,GHz.  Contours in
  each frame denote the 5$\sigma$ significance contours on the
  870\um\ image (also shown in Figure~\ref{fig:maps}); 
    the white dotted line shows the aperture (based on IRAC emission)
    used to measure photometry in the optical and near-infrared.  The
  only significant ($>$3$\sigma$) detections come from
  {\it Spitzer} IRAC, ALMA, and VLA at 3\,GHz. Below, the aggregate
  composite SED for both components A+B is shown in black, made of
  three primary components: the stellar and nebular line emission
  (dark blue), the thermal dust emission (orange), and synchrotron
  radio emission (purple).  All three components are independently fit
  to data in their respective regimes.  The light blue curve shows the
  modeled unattenuated stellar and nebular emission (described in
  text); the dotted orange line shows the dust SED fit to component A,
  the dashed orange line is for component B, and the gray line shows
  the best-fit {\sc MAGPHYS} SED.  Our SED does not include emission
  from PAHs in the mid-infrared due to the existing dearth of data in
  that regime.  Upper limits are shown as 2$\sigma$.  The inset plot
  shows the probability distributions of stellar mass derived for
  \source\ from the OIR-only fit (blue) and {\sc Magphys} fit (gray).}
\label{fig:fullsed}
\end{figure*}

Because the size measurements for component A and component B are
broadly consistent, and we lack data for a more detailed analysis, we
use the measured 870\um\ dust-emitting sizes to estimate the galaxies'
dynamical masses.  There are a number of potential caveats in doing
this. First, the dynamical mass is best measured in the same tracer
used to infer the galaxy's kinematics (\cof\ in this case).  Second,
using a high-J tracer like \cof\ would likely bias the dynamical mass
estimate because it only probes dense gas regions.  While both of
these concerns are important to keep in mind, a few facts provide
reassurance that our assumptions are sufficient in this case: first,
the fact that the galaxy's 3\,mm size is not significantly larger than
its 870\um\ size, and second, the fact that the uncertainty on the
 measured quantities dominates the
calculation of the dynamical mass. In other words, the
  uncertainty on $R_{\rm e}$ and $\sigma$, combined with uncertainty
  in unconstrained $i$ are significant enough to dominate over
  variations in tracer-dependent forms of these quantities.

Thus, we adopt circularized effective radii of $R_{\rm
  e}(A)=380\pm30$\,pc and $R_{\rm e}(B)=760\pm130$\,pc.  The velocity
dispersions as measured from \cof\ (and as shown in
Figure~\ref{fig:specdetail}) are $\sigma_{\rm
  V}(A)=$\,260$\pm$40\,\kms\ and $\sigma_{\rm
  V}(B)=$\,280$\pm$130\,\kms.  This gives dynamical mass estimates of
$M_{\rm dyn}(A)=\,$(5$\pm$2)$\times10^{10}$\,\msun\ and $M_{\rm
  dyn}(B)=$(7$\pm$6)$\times10^{10}$\,\msun, respectively.  Though the
dynamical mass estimated for component B is a bit larger (due to its
larger physical size) the uncertainty is quite large and consistent
with being an equal or smaller mass companion.  While the mass
calculated for component A seems rather precise, it should be noted
that the double-peaked \cof\ spectrum of component A is poorly fit to
a single Gaussian component.  Note that if we instead
  calculate a dynamical mass from the \htwoo\ line width, which is
  much more uncertain, we get dynamical mass estimates an order of
  magnitude larger; as Figure~\ref{fig:specdetail} shows, this is not
  because the \htwoo\ line is much more broad than the \cof\ line, but
  it represents the difference between a single and double component
  fit.

Alternatively, the dynamical mass of component A could be estimated
directly from the resolved \cof\ kinematics using
\begin{equation}
M_{\rm dyn}=\frac{V_{\rm
  max}^{2}\,R_{\rm max}}{G}
\end{equation}
which then similarly needs to be corrected for unknown inclination.
Note that $R_{\rm max}$ here denotes the maximum radius at which
$V_{\rm max}$ is measured and differs from $R_{\rm e}$, the
circularized half-light radius, used above.  Using both size $R_{\rm
  max}=$\,0\farcs85$\pm$0\farcs20 = 5.0$\pm$1.2\,kpc and $V_{\rm
  max}=$\,350$\pm$50\,\kms\ measurements from the position-velocity
diagram in Figure~\ref{fig:kinematics}, we derive an alternate
dynamical mass of component A of (2.0$\pm$0.8)$\times10^{11}$\,\msun.
 The dynamical mass for component B cannot be constrained
  using this method because of the low signal-to-noise of the line and
  no resolved rotation curve.

Given the caveats of using a different tracer for size measurements,
we adopt the more conservative higher-mass dynamical constraint for
component A of (2.0$\pm$0.8)$\times10^{11}$\,\msun, but discuss the
implications of either dynamical mass constraint in our discussion of
the total mass budget in \S~\ref{sec:massbudget}.  For component B, we
adopt the only estimated, yet highly uncertain, value of $M_{\rm dyn}$
of (7$\pm$6)$\times10^{10}$\,\msun.

\subsection{SED Fitting}

We fit spectral energy distributions using three methods: the stellar
component only \citep[as in][]{finkelstein15a}, the obscured component
only \citep[as in][]{casey12a}, and both together using energy balance
techniques \citep[specifically {\sc
    magphys};][]{da-cunha08a,da-cunha15a}.  The three approaches, used
to derive different physical quantities, are described below. The
derived properties are given in
Table~\ref{tab:physical}.  As Figure~\ref{fig:fullsed}
  shows, the final SED we adopt for \source\ is outlined in black; the
  details are described throughout this section.

\begin{table*}
\caption{Derived Properties of the \source\ System.}
\centering
\begin{tabular}{lcccc}
\hline\hline
Derived &  Units & Component & Component & Total \\
Property &  & A & B & A+B \\
\hline
RA & --- & 10:00:26.356 & 10:00:26.356 & --- \\
DEC & --- & $+$02:15:27.94 & $+$02:15:26.63 & --- \\
\multicolumn{5}{c}{\underline{\it From ALMA Spectroscopy:}}\\
$z$ & --- & 5.850 & 5.852 & 5.850 \\
I$_{\rm CO(6-5)}$ & Jy\,km\,s$^{-1}$ & 0.43$\pm$0.03 & 0.07$\pm$0.02 & 0.48$\pm$0.03 \\
$\sigma_v(CO)$ & \kms\ & 260$\pm$40$^{a}$ & 280$\pm$130 & 700$\pm$70 \\
L$_{CO(6-5)}^\prime$ & K\,km\,s$^{-1}$\,pc$^{2}$ & (1.4$\pm$0.9)$\times10^{10}$ & (2.3$\pm$0.7)$\times10^{9}$ & (1.56$\pm$0.10)$\times10^{10}$ \\
I$_{\rm H_{2}O(2_{1,1}-2_{0,2})}$ & Jy\,km\,s$^{-1}$ & 0.09$\pm$0.02 & --- & 0.09$\pm$0.02 \\
$\sigma_{v}(H_{2}O)$ & \kms\ & 900$\pm$200 & --- & 900$\pm$200 \\
L$_{\rm H_{2}O(2_{1,1}-2_{0,2})}^\prime$ & K\,km\,s$^{-1}$\,pc$^{2}$ & (2.5$\pm$0.5)$\times10^{9}$ & --- & (2.5$\pm$0.5)$\times10^{9}$ \\
M$_{\rm gas}({\rm CO})$ & \msun\ & (3.3$\pm$1.7)$\times10^{11}$ & (5$\pm$3)$\times10^{10}$ & (3.7$\pm$1.8)$\times10^{11}$ \\
M$_{\rm dyn}$ & \msun & $(2.0\pm0.8)\times10^{11}$ & $(7\pm6)\times10^{10}$ & --- \\
\multicolumn{5}{c}{\underline{\it From ALMA Dust Continuum:}}\\

M$_{\rm dust}$ & \msun\ & (1.3$\pm$0.3)$\times$10$^{9}$ & (1.9$^{+1.3}_{-0.8}$)$\times10^{8}$ & (1.6$^{+0.4}_{-0.3}$)$\times10^{9}$ \\
M$_{\rm gas}({\rm 3\,mm})$ & \msun & (1.4$\pm$0.4)$\times$10$^{11}$ & (1.2$\pm$0.5)$\times10^{10}$ & (1.7$\pm$0.4)$\times10^{11}$ \\
FWHM$_{\rm maj}$(870\um)$^{b}$ & $''$ & 0\farcs15$\pm$0\farcs01 & 0\farcs30$\pm$0\farcs05 & --- \\
Axis Ratio (870\um)$^{b}$ $b/a$ & --- & 0.87$\pm$0.15 & 0.37$\pm$0.24 & --- \\
$R_{\rm eff}$($870$\um) & pc & 380$\pm$30 & 760$\pm$130 & --- \\
FWHM$_{\rm maj}$(3\,mm)$^{b}$ & $''$ & 0\farcs29$\pm$0\farcs11 & --- & --- \\
$R_{\rm eff}$(3\,mm) & pc & 700$\pm$300 & --- & --- \\
$\Sigma_{M_{\rm dust}}$ & \msun\,pc$^{-2}$ & 1400$\pm$400 & 50$\pm$30 & --- \\
$\Sigma_{\rm SFR}$ & \msun\,yr$^{-1}$\,kpc$^{-2}$ & 640$\pm$170 & 61$\pm$35 & --- \\
\multicolumn{5}{c}{\underline{\it From Broad-band SED Fitting:}}\\
$\lambda_{\rm rest}$ at which $\tau_{\nu}=1$ & \um\ & $\equiv$200 & Opt. Thin & --- \\
\lir & \lsun & (4.0$^{+0.9}_{-0.7}$)$\times10^{12}$ & (1.5$^{+1.1}_{-0.5}$)$\times10^{12}$ & (6.3$^{+1.1}_{-0.9}$)$\times10^{12}$ \\
SFR & \sfr & 590$^{+140}_{-100}$ & 220$^{+150}_{-70}$ & 930$^{+160}_{-130}$ \\
\lpeak & \um & 87$\pm$7 & 100$^{+30}_{-80}$ & --- \\
T$_{\rm dust}$ & K & 56.3$^{+5.9}_{-5.7}$ & 29.7$^{+8.5}_{-6.6}$ & --- \\
$\beta$ & --- & 1.95$\pm$0.11 & 2.66$^{+0.22}_{-0.34}$ & --- \\
M$_{\star}$ ({\sc OIR-only})& \msun & --- & --- & (3.2$^{+1.0}_{-1.5}$)$\times10^{9}$ \\
L$_{\rm UV}$(1600\AA) & \lsun & $<$7.7$\times10^{9}$ & (8.8$\pm$3.6)$\times10^{9}$ & $<$3.4$\times$10$^{10}$ \\
IRX & --- & $>$510 & 160$^{+280}_{-100}$ & --- \\
A$_{\rm UV}$ & --- & $>$6.2 & 5.0$^{+1.0}_{-1.1}$ & --- \\
$q_{\rm IR}$ & --- & 0.4$\pm$1.0 & --- & --- \\
M$_{\rm halo}$ & \msun & --- & --- & (3.3$\pm$0.8)$\times$10$^{12}$ \\
\hline\hline
\end{tabular}
\label{tab:physical}

\begin{minipage}{\textwidth}
{\small {\bf Notes.}  Positions measured from
    870\um\ dust continuum.  Note that quantities derived for the
    total \source\ system (A+B) are derived independently from the
    measurements of the two individual systems.  In other words, Total
    is not simply the sum of the two, but a direct independent
    measurement of the integrated quantity.
A brief guide to derived properties: 
I$_{\rm CO(6-5)}$ is the integrated line flux of the \cof\ line,
$\sigma_{v}(CO)$ is the velocity dispersion of the \cof\ feature,
$L^\prime_{\rm CO(6-5)}$ is the \cof\ line luminosity.  All three
quantities are similarly derived for the \htwoo\ line.
M$_{\rm gas}(CO)$ is the gas mass as derived from the \cof\ line while
M$_{\rm gas}(3\,mm)$ is the gas mass as derived from 3\,mm dust
continuum (and M$_{\rm dust}$ is the dust mass derived from 3\,mm
continuum).  Both assume $\alpha_{\rm
  CO}\,=$\,6.5\,\msun\,(K\,km\,s$^{-1}$\,pc$^{2}$)$^{-1}$.
M$_{\rm dyn}$ is the dynamical mass as estimated from the \cof\ line
width and 870\um\ dust continuum size.
FWHM$_{\rm maj}$ is the measured full-width at half maximum size
measured from dust continuum images at 870\um\ or 3\,mm (in the image
plane), and the axis ratio indicates the relative elongation on the
plane of the sky.  R$_{\rm eff}$ is the circularized half-light radius
in parsecs.  $\Sigma_{\rm M_{\rm dust}}$ and $\Sigma_{\rm SFR}$ are
the dust mass surface density and star formation surface density,
respectively.
$\lambda_{\rm rest}$ is the wavelength at which $\tau=1$ for the dust
SED, \lir\ is the derived IR luminosity integrated from 8--1000\um,
and SFR is the star-formation rate converted directly from \lir\ using
the \citet{kennicutt12a} scaling.  $\lambda_{\rm peak}$ is the
rest-frame wavelength where the dust SED peaks, and $T_{\rm dust}$ is
the underlying dust temperature used in the model fit to the
photometry.  $\beta$ is the emissivity spectral index.  $M_{\star}$is
the stellar mass of the aggregate system, while $L_{\rm UV}$ is the
rest-frame 1600\AA\ luminosity.  IRX is the ratio of \lir/$L_{\rm
  UV}$, A$_{\rm UV}$ is the absolute magnitudes of attenuation
inferred at 1600\AA, $q_{\rm IR}$ is the implied FIR-to-radio ratio as
in \citet{yun01a}, and $M_{\rm halo}$ is the total inferred halo mass.
$^{a}$ Component A is best characterized by
double-peaked emission for which the line width of each component has
width of 370\,\kms.
$^{b}$ FWHM of the major axis measured from a two-dimensional Gaussian
fit in the image plane, and the axis ratio is from the same fit.}
\end{minipage}
\end{table*}

\subsubsection{Optical/Near-Infrared SED Fit}\label{sec:oir}

We explore what constraints can be set using the optical and
mid-infrared photometry alone. As the stellar component is detected
only in the deepest near-infrared imaging, we use only the {\it HST}
imaging from COSMOS \citep{scoville07a} and CANDELS
\citep{koekemoer11a,grogin11a} data, in addition to S-CANDELS {\it
  Spitzer} IRAC measurements \citep{ashby15a} for the
optical/near-infrared fit; all other existing data is not deep enough
to provide meaningful constraints.

We measure photometry using Source Extractor \citep{bertin96a}, using
a combined [3.6]+[4.5] image as the detection image.  We optimize the
detection parameters such that the isophotal region corresponds to an
ellipse which includes the majority of the bright IRAC emission, and
is tuned to enclosed both ALMA 870\um\ continuum peaks (see
Figure~\ref{fig:maps}).  We up-sample the IRAC images to the same
0.06$''$ pixel scale as the {\it HST} photometry (altering the
zero-point appropriately), and run Source Extractor with the {\it HST}
F606W, F814W, F125W, F140W and F160W images as the measurement images.
Though this area is covered by shallow F140W from the 3D-HST Survey,
\source\ falls in a coverage gap \citep{momcheva16a}.  
As expected, we find
a significant detection in both IRAC bands, with no significant flux
measured in any of the {\it HST} bands.

We use this isophotal photometry to estimate the stellar population
modeling parameters following the methodology of
\citet{finkelstein15a}.  In brief, we use the EAZY software
\citep{brammer08a} to fit the photometry using the updated templates
derived from Flexible Stellar Population Synthesis (FSPS) models
\citep[][see a forthcoming paper by S. Finkelstein for more details on
  the templates)]{conroy09a,conroy10a}.  In the absence of a
spectroscopic identification, such little photometric information
would result in a photo-z probability distribution that is very broad,
consistent with $z > 2$.  We then performed SED fitting using the
spectroscopic redshift performing $\chi^2$ minimization of a set of
\citet{bruzual03a} stellar population models, with added nebular
emission and dust attenuation.  The results are illustrated in the
optical portion of the full SED shown in Figure~\ref{fig:fullsed}
(blue lines).  The inset plot shows the inferred distribution of
stellar mass for the best-fit `OIR' SEDs (in blue), with a median of
(3.2$^{+1.0}_{-1.5}$)$\times$10$^{9}$\,\msun.  From the limited
optical/near-infrared data alone, the absolute magnitudes of
attenuation estimated in the rest-frame $V$ band is $A_{V}=3.1\pm0.2$
and SFR = 63.4$^{+6.5}_{-6.8}$\,\sfr\ (corrected for `dust' that is
estimated from the OIR fit). Both are underestimated relative to the
measured characteristics of the far-infrared SED.

The inferred UV luminosity from the best-fit SED is extrapolated to be
$L_{\rm 1600}\approx7\times10^{8}$\,\lsun, though observationally it
is only strictly limited to $L_{\rm 1800}\simlt7\times10^{10}$\,\lsun
(at 2$\sigma$, based on the F125W non-detection).  To set more
stringent constraints for the individual components A and B, we
extract 0.6$''$ circular aperture photometry on the {\it HST} data.
While component A is not detected, there is a 2$\sigma$ marginal
detection in component B.  We use these measured UV constraints to
also constrain IRX, defined as the ratio of $L_{\rm IR}$/$L_{\rm UV}$,
and the absolute magnitudes of attenuation in the UV, A$_{\rm UV}$, in
the samples.  These measurements use the
  \lir\ calculated in the next section, \S~\ref{sec:irsed}, and given
  in Table~\ref{tab:physical}. For component A, we measure
IRX\,$>$\,510 and $A_{\rm UV}>6.2$, while for component B we measure
IRX\,=\,160$^{+280}_{-100}$ and $A_{\rm UV}=5.0^{+1.0}_{-1.1}$.  Even
though the difference between the two components may be substantial,
both constitute extremely obscured systems.

\subsubsection{Far-IR/millimeter SED Fit}\label{sec:irsed}

The obscured SED (probed by rest-frame wavelengths $\sim$5--3000\um)
has no detections at wavelength shortward of 850\um.  Nevertheless,
due to the superb quality of the ALMA continuum detections, we can fit
a somewhat well-constrained obscured SED with a single modified
blackbody plus mid-infrared powerlaw.  The mid-infrared component is
unconstrained due to the dearth of detections shortward of the peak,
but we adopt a model that follows $S_{\nu}\propto\nu^{-\alpha}$
\citep[the powerlaw joins the modified blackbody at the point where
  the derivative is equal to $\alpha$, as described in][]{casey12a}.
Here we fix the value of $\alpha$ to $\alpha_{\rm MIR}=4$.
Physically, a lower value of $\alpha$ represents a higher proportion
of emission emanating from dust heated from discrete sources rather
than the cooler dust heated by the ambient radiation field in the ISM.
Very high values of $\alpha$ asymptote to the pure modified blackbody
solution.  While no direct constraints can be made for
  $\alpha_{\rm MIR}$, it should be noted that values less than
  $\alpha_{\rm MIR}=4$ violate the upper limits in the mid-infrared as
  measured by {\it Spitzer} and {\it Herschel}.  If $\alpha_{\rm MIR}$
  is constrained to values in excess of $\sim$4, then the total impact
  on the integrated \lir\ is negligible.
The SED is fit using a simple Metropolis Hastings Markov
Chain Monte Carlo with free parameters of \lpeak, the rest-frame peak
wavelength, \lir, the integrated 8--1000\um\ IR luminosity, and
$\beta$, the emissivity spectral index.  This is an updated fitting
technique that largely follows the methodology outlined in
\citet{casey12a} but substitutes least squares fitting for Bayesian
analysis and a contiguous function for a piecewise powerlaw and
modified blackbody (this will be described in a forthcoming paper by
P. Drew). This fit embeds the impact of CMB heating on ISM dust at
high-redshift as prescribed in \citet{da-cunha13a}; at $z=5.85$, the
CMB is 18.7\,K.  Note that \lpeak, \lir, and $\beta$ (the three free
parameters of the fit) are measured from the {\it intrinsic} emitted
SED, not the observed SED and observed photometry which has been
impacted by the CMB.

\begin{figure*}
\centering
\includegraphics[width=1.99\columnwidth]{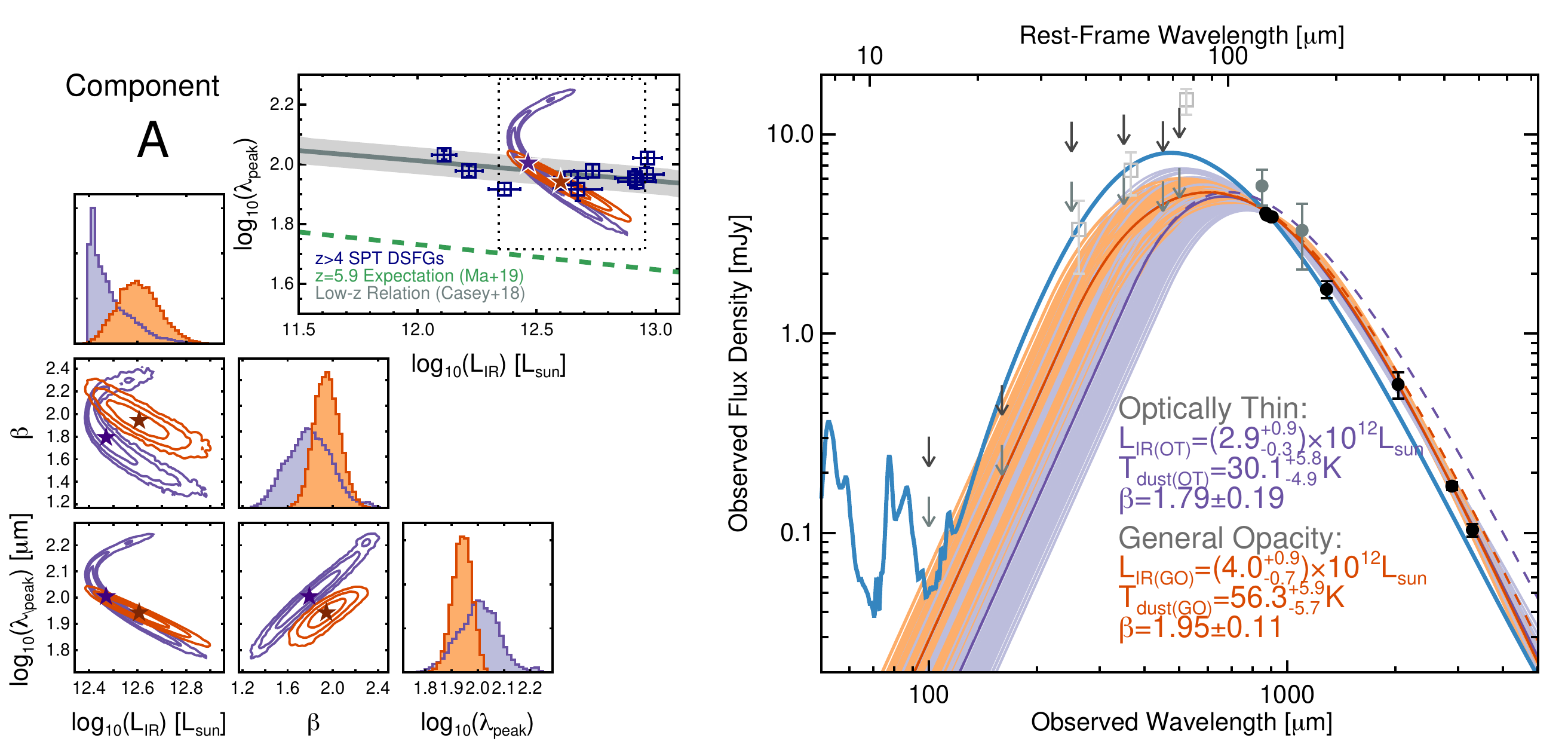}
\includegraphics[width=1.99\columnwidth]{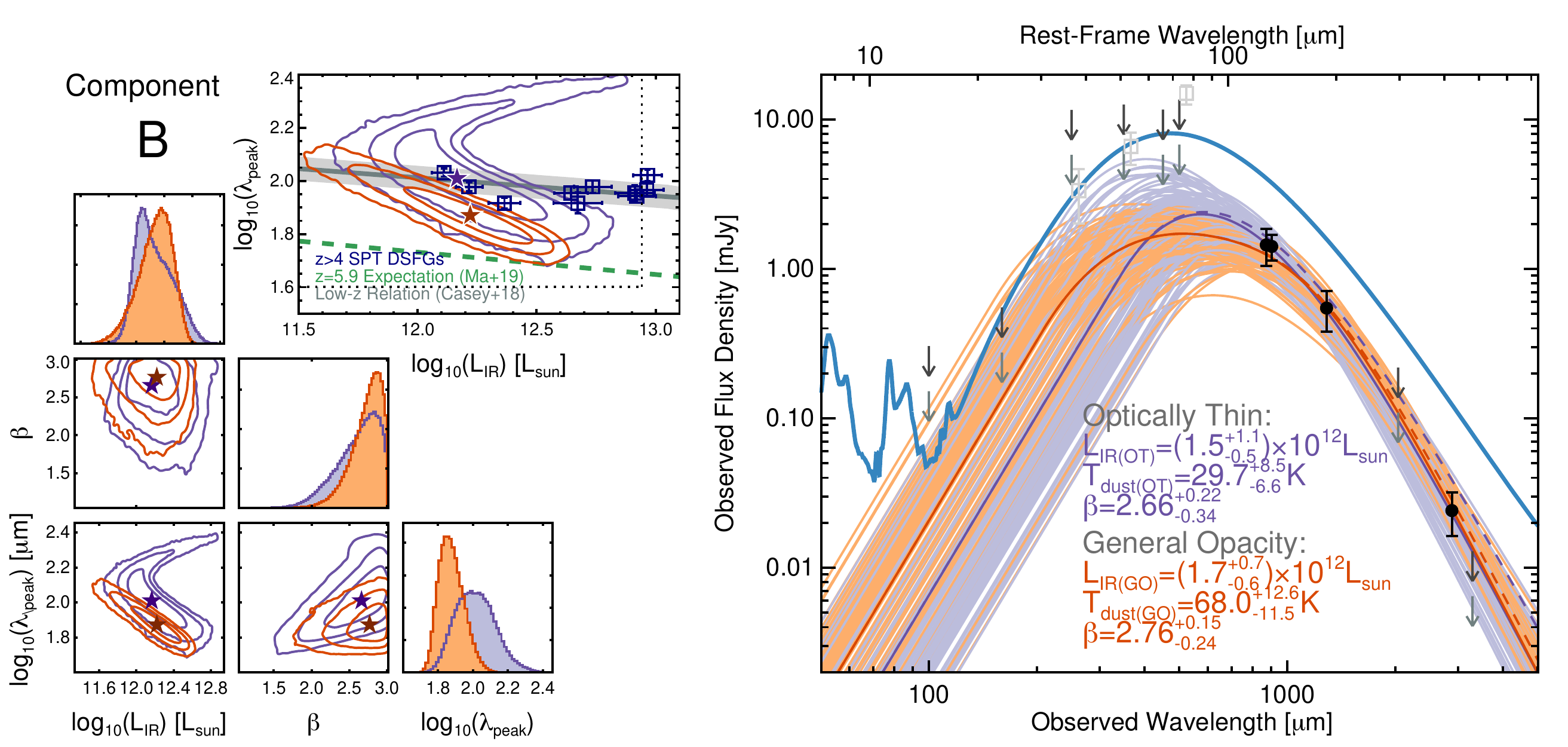}
\caption{Far-infrared SED fitting details for both components of
  \source, adopting two different model assumptions: optically thin
  dust (purple), and a more general opacity model (orange) that
  asserts $\tau=1$ at a rest-frame of 1.5\,THz (200\,\um). Corner
  plots are shown for the converged MCMC chains at left constraining
  \lir, \lpeak, and $\beta$ for each component.  The upper middle
  panels show \lir-\lpeak\ for each source in context against
  expectation from cosmological simulations \citep[green dashed
    line][]{ma19a} as well as against the average \lir-\lpeak\ derived
  for lower-redshift DSFGs \citep[gray band;][]{casey18a}.  DSFGs at
  $z>4$ from the SPT survey are shown as navy squares
  \citep{strandet16a}.  The lack of detections shortward of the peak
  imply that the mid-infrared powerlaw cannot be directly constrained
  and here we fix $\alpha_{\rm MIR}=4$, which minimally impacts the
  measured \lir.  Detections above 5$\sigma$ significance are shown in
  black, $2.5\sigma<S/N<5\sigma$ detections in gray, and 2$\sigma$ and
  3$\sigma$ upper limits as light and dark gray arrows.  The
  ``super-deblended'' {\it Herschel} photometry shown in
  \citet{jin19a} is shown in light open squares but are not used for
  these fits.   The best-fit {\sc Magphys} fit is shown
    in a thick blue line, while random draws from the accepted MCMC
    trials are shown in either light orange (general opacity) or
    purple (optically thin), with the median value SEDs shown in dark
    orange or purple. At $z=5.85$ the impact of the CMB on the SED is
  model dependent: the intrinsic emitted SEDs we would expect in the
  absence of the CMB are shown as dashed purple and orange lines,
  measurably different only for the optically thin model.}
\label{fig:irsed}
\end{figure*}

Figure~\ref{fig:irsed} shows the results of the obscured SED fit for
both optically thin and a more general opacity model for the two major
components of this source.  The general opacity model assumes $\tau=1$
at rest-frame 1.5\,THz \citep[or 200\um;][]{conley11a}.  Due to the
high signal-to-noise ratio of the ALMA measurements, in particular the
870\um\ data near the SED peak, the long wavelength portion of the SED
and the peak are more precisely constrained than for most DSFGs and
also allows for an independent measurement of $\beta$, the emissivity
spectral index.  For component B, we set an upper limit to $\beta=3$
based on the low S/N of the source's photometry.
The lower left section of each component fit shows a corner plot of
the converged MCMC chains in \lpeak, \lir, and $\beta$.  Both
optically thin and general opacity models are significantly higher
quality for component A than component B.  The upper middle panel
places the measured \lir\ and \lpeak\ values in context against: (a)
the $z\sim1-2$ \lir-\lpeak\ relationship derived for DSFGs in
\citet{casey18a}, (b) the measured characteristics of $z>4$ DSFGs from
the SPT survey \citep{strandet16a}, and (c) in contrast to expectation
for $z\sim6$ galaxies from theoretical modeling \citep{ma19a}.  Note
that while several modeling papers
\citep[e.g.][]{behrens18a,ma19a,liang19a} suggest that the
luminosity-weighted dust temperatures of galaxies at $z\simgt5$ should
be warmer than those at $z\sim2$, we do not see compelling evidence
that this holds for either component of \source.

The parameter \lpeak\ is preferred over a direct fit to the physical
dust temperature because it is insensitive to the opacity model
assumed; in other words, an SED that peaks at rest-frame 90\um\ could
have an intrinsic dust temperature ranging anywhere from 30--50\,K
depending on the geometry and column density of the dust.  But because
\source\ sits at such a high redshift where CMB heating is
non-negligible, the opacity model assumptions do impact the
intrinsic rest-frame peak wavelength \lpeak.  Fit to the same
photometry, optically thin dust SEDs will consistently have lower dust
temperatures than more general opacity assumptions allowing for dust
self-absorption near the peak, thus they are proportionally more
impacted by CMB heating. 
Thus, the difference in measured \lpeak\ in Figure~\ref{fig:irsed}
between opacity models is purely due to the different levels of impact
of the CMB.

While the CMB does impact \lpeak\ by way of the underlying physical
dust temperature, the gap between \lir\ that is fit with different
opacity models is smaller than what it would be in the absence of the
CMB or at lower redshifts.  As shown on the right-hand panels of
Figure~\ref{fig:irsed}, the difference between the emitted SED (dashed
line) and observed SED (dark solid line) is much larger for the
optically-thin SED (purple) than for the general opacity model
(orange), so while the CMB has little impact on the derived \lir\ of
the general opacity model fits, it has small but measurable impact on
\lir\ of the optically-thin model.  

Through analysis of the dust mass surface density from the
870\um\ data, we have roughly constrained the wavelength at which the
SED becomes optically thick.  $\Sigma_{\rm M_{\rm
    dust}}=1400\pm400$\,\msun\,pc$^{-2}$ in component A suggest an
optically thick SED to $\sim$200--300\um\ rest-frame,
while the lower dust column density in component B of $\Sigma_{\rm
  M_{\rm dust}}=50\pm30$\,\msun\,pc$^{-2}$ suggest the SED is
optically-thin near the peak \citep[these measurements assume the dust
  mass absorption coefficients as given in][]{li01a}.  These different
dust column densities imply that the dust SEDs of the two components
should be treated differently, with component A more reminiscent of
the very high column densities of dust that is ubiquitous among the
brightest DSFGs at lower redshifts.  Thus, we adopt the more general
opacity model (with $\tau=1$ at $\lambda_{\rm rest}=200$\,\um) for
component A and the optically-thin model for component B.

The implied SFRs from the dust emission, assuming the
\citet{kennicutt12a} scaling \citep[which uses an IMF
  from][]{kroupa03a}, are 590$^{+140}_{-100}$\,\sfr\ for component A
and 220$^{+150}_{-70}$\,\sfr\ for component B.  Note that the total
SFR fit to the aggregate SED (930$^{+160}_{-130}$\,\sfr) is lower
than, though not fully inconsistent with, the derived SFR of the
system in \citet{jin19a} of 1283$\pm$173\,\sfr.  While the dust
temperature of the general opacity model may seem high compared to
some DSFGs in the literature, Figure~\ref{fig:irsed} shows they are
fully consistent with the measured \lpeak\ values for both
lower-redshift DSFGs \citep[as derived in][]{casey18a} as well as
existing measurements for $z>4$ DSFGs from the SPT survey
\citep{strandet16a}.  They are both colder (i.e. higher $\lambda_{\rm
  peak}$) than the expected median \lir-\lpeak\ relation from
simulations at $z\sim5.9$ \citep{ma19a}. 

There is a slight discrepancy between measured emissivity spectral
index, $\beta$, of components A and B of $\beta(A)=1.95\pm0.11$ and
$\beta(B)=2.66^{+0.22}_{-0.34}$; while this could be a real
discrepancy, the quality of the constraint for component B is weak,
and there is a significant degeneracy between $\beta$ and
$\lambda_{\rm peak}$ in the absence of high signal-to-noise data on
the Rayleigh-Jeans tail.  Note that \citet{jin19a} conclude that the
CMB might be impacting the net SED by artificially steepening $\beta$
for \source, however we do not see evidence for this in the case of
component A and we only see weak evidence that this is the case for
component B.  In other words, if we fit the SED of component B without
accounting for CMB heating, we would measure a steeper value of
$\beta=2.89\pm0.40$ than we do having accounted for the CMB.
The difference in our conclusions regarding component A is driven by
the inclusion (or not) of the single dish photometry from {\it
  Herschel} (in particular the deblended photometry), {\sc AzTEC}, and
{\sc SCUBA-2}.  The band 7 data presented in this paper results in a
much less steep Rayleigh-Jeans tail and derived value of $\beta$
consistent with the often-used assumption in the literature of
$\beta=1.5-2.0$.

In conclusion, we infer that component A is optically
  thick at the peak, has a measured
  \lir=(4.0$^{+0.9}_{-0.7}$)$\times10^{12}$\,\lsun,
  SFR=590$^{+140}_{-100}$\,\sfr, rest-frame peak wavelength of
  $\lambda_{\rm peak}$=87$\pm$7\,\um, dust temperature $T_{\rm
    dust}=$56.3$^{+5.9}_{-5.7}$\,K, and emissivity spectral index
  $\beta$=1.95$\pm$0.11.  We infer that component B is optically thin
  at the peak, has a measured
  \lir=(1.5$^{+1.1}_{-0.5}$)$\times10^{12}$\,\lsun,
  SFR=220$^{+150}_{-70}$\,\sfr, rest-frame peak wavelength
  $\lambda_{\rm peak}$=100$^{+30}_{-80}$\,\um, dust temperature
  $T_{\rm dust}$=29.7$^{+8.5}_{-6.6}$\,K, and emissivity spectral
  index $\beta$=2.66$^{+0.22}_{-0.34}$.

\subsubsection{Energy Balance SED Fit}

We employ the updated library of star-formation histories described in
\citet{da-cunha15a} and stellar population synthesis models from
\citet{bruzual03a} to fit the full SED (both components together)
using {\sc Magphys}.  The advantage of {\sc Magphys} comes through the
use of energy-balance, whereby energy absorbed in the rest-frame UV
and optical is re-radiated by dust at long wavelengths.  However, this
technique can break down for sources whose stellar emission is fully
decoupled from long-wavelength emission, as is often the case in DSFGs
\citep[e.g.][]{casey17a}.

We fit both components of \source\ (A+B) jointly using {\sc Magphys}
due to the difficulty of differentiating IRAC fluxes.  The best-fit
{\sc Magphys} solution is shown in Figure~\ref{fig:fullsed} as well as
Figure~\ref{fig:irsed} for comparison against our adopted best-fit
OIR-only and obscured-only SED.  The {\sc Magphys} fit does well
fitting the total IR photometry and upper limits in the mid-infrared
and, likewise, provides a sensible solution to the limited OIR
photometry.  The probability distribution of stellar mass as derived
by {\sc Magphys} is shown as the inset plot on
Figure~\ref{fig:fullsed} (in gray), with
$M_{\star}=(2.1\pm0.7)\times10^{11}$\,\msun.  The predicted $A_{\rm
  V}$=5.64$^{+0.02}_{-0.20}$ is well-aligned with the measured
constraint on $A_{\rm UV}$.

There is a marked difference between this stellar mass estimate and
that provided by the analysis of the fit to the OIR component only,
roughly a factor of $\sim$60$\times$ different, with estimated
uncertainties far smaller than the relative offset.  This can largely
be attributed to the lack of constraint near the rest-frame
1.6\um\ stellar bump (redshifted to $\sim$11\um). The SEDs in
Figure~\ref{fig:fullsed} show a stark difference near 10\um\ which
results in these disparate predictions.  Whether or not nebular
emission lines are included in the stellar population model also
impacts the results but to a lesser degree. Several
  works have shown that high equivalent-width emission lines can
  enhance broadband flux in the IRAC channels by a factor of
  $\sim$2--3$\times$ \citep{shim11a,salmon15a}. The OIR-only SED
fitting from \S~\ref{sec:oir} included them, which would potentially
drive the stellar mass estimate lower for a given set of photometry,
the stellar populations models used by {\sc Magphys} do not include
them.  At this redshift, H$\alpha$\,6563\AA\ emission falls into the
[4.5] IRAC band, while OIII\,5007\AA\ falls into the [3.6] IRAC band,
and sources with significant star-formation and ionization radiation
will be proportionally brighter in these bands as a result
\citep[e.g.][]{chary05a,smit14a}, if those lines are not substantially
dust-obscured themselves \citep[which is often the case for
  DSFGs;][]{swinbank04a,casey17a}.

A detailed analysis of the stellar masses of submillimeter galaxies
was performed by \citet{michaowski14a}, who used synthetic photometry
of simulated DSFGs to test different literature tools used to infer
their stellar mass (among other characteristics).  They found that
tools like {\sc Magphys} that model two independent star-formation
histories (continuous and starburst) best reproduce the simulated DSFG
stellar masses while single-component star-formation history models
tend to underestimate the stellar mass substantially.  A comparison of
the two stellar mass estimates to other measured quantities in the
system, like gas mass and dynamical mass, also suggest that the {\sc
  Magphys}-predicted stellar mass may be the more likely of the two
(i.e. roughly equal stellar masses and gas mass, rather than a gas
mass than exceeds the stellar mass by $\sim$60$\times$).

After an analysis of the galaxy's total mass budget
and predicted halo mass, given in \S~\ref{sec:discussion}, we find
that the stellar mass is more likely consistent with the OIR-only
value,  though formally unconstrained, with the
  possibility of being within the very large range of a few
  $\times$10$^{9}$\,\msun\ to a few $\times10^{11}$\,\msun.

\subsection{Radio Continuum Emission}

In the radio regime, \source\ component A is marginally detected at
$3.2\sigma$ significance in the very deep 3\,GHz map
\citep{smolcic17a} and not detected at 1.4\,GHz \citep{schinnerer07a}.
In the absence of direct constraints on the radio SED, we adopt a
synchrotron slope of $\alpha_{\rm rad}=-0.8$ \citep[
    consistent with][ often used in the literature when
    there is a dearth of multiple radio continuum
    constraints]{condon92a} fit to the 3\,GHz photometry.  The
far-infrared/radio correlation for star-formation would predict radio
emission below the detection limit in the case of both a
non-evolving relation \citep[$\sim$50\,nJy, expected $q_{\rm
    IR}=2.4$;][]{yun01a} and evolving relation \citep[$\sim$300\,nJy,
  expected $q_{\rm IR}=2.0$;][]{delhaize17a}.  The marginal 3\,GHz
detection implies a value of $q_{\rm IR}=0.4\pm1.0$, suggestive of a
possible buried AGN.

 Though suggestive of an AGN, it should be noted that the
  {\sc Magphys} SED, which assumes a non-evolving FIR-radio
  correlation $q_{\rm IR}=2.34$ \citep{da-cunha15a}, intersects our
  measured 3\,GHz radio continuum measurement without need to invoke a
  possible radio AGN.  This is due to a higher \lir\ for the {\sc
    Magphys} fit, largely driven by excess emission in the
  mid-infrared regime, where we lack data.  It is also due to the
  assumed synchrotron slope in the radio regime, closer to
  $\alpha=-0.65$, and incorporating emission from free-free emission.
  The combination of these different assumptions lead to a
  star-formation-dominated radio SED in line with measurements.

Whether or not there is an AGN in \source\ is unclear.
  It is not detected in the {\it Chandra} X-ray imaging in COSMOS,
though no detection would be expected at the given depth for a
$z=5.85$ buried AGN.  Future observations of the CO spectral line
energy distribution, other (sub)millimeter emission features, as well
as rest-frame optical emission lines from {\it JWST} will play a
crucial role in inferring whether or not such a buried AGN exists.

\section{Discussion}\label{sec:discussion}

\subsection{Mass Budget and Halo Mass Rarity}\label{sec:massbudget}

\begin{figure}
\includegraphics[width=0.99\columnwidth]{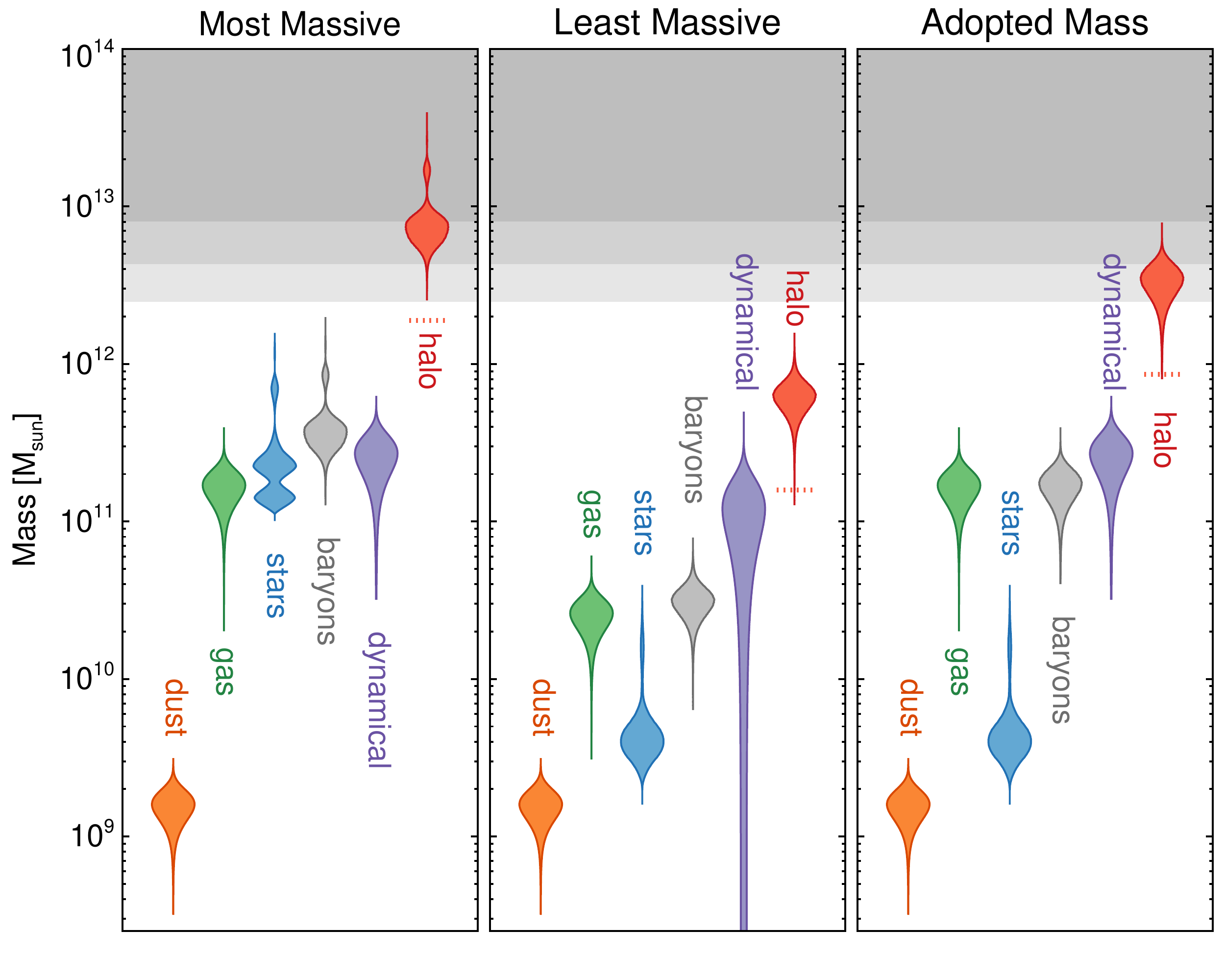}
\caption{The mass budget of the \source\ system, considering both the
  most massive assumptions (left), the least massive assumptions
  (middle) and adopted masses (right).  This violin plot
    shows the probability density distribution on its side for each
    measured variable.  The most massive assumptions are derived
  using $\alpha_{\rm
    CO}=6.5$\,\msun\,(K\,km\,s$^{-1}$\,pc$^{2}$)$^{-1}$ for the gas
  mass, the {\sc Magphys}-derived stellar mass, and the
  kinematically-derived dynamical mass from
  Figure~\ref{fig:kinematics}.  The least massive assumptions are
  derived using $\alpha_{\rm
    CO}=1$\,\msun\,(K\,km\,s$^{-1}$\,pc$^{2}$)$^{-1}$ for the gas
  mass, the OIR-only derived stellar mass, and the unresolved
  dynamical mass estimate.  The dust mass is the same in all cases.
  The halo mass is extrapolated from the total baryonic mass assuming
  a conservative 1:20 baryonic-to-halo mass ratio; the lower limit on
  halo mass is the dotted red line, set by the cosmic baryon fraction.
  The gray shaded regions represent the 1$\sigma$, 2$\sigma$, and
  3$\sigma$ exclusion curves for finding a galaxy above a given mass
  at $z=5.85$ in our 2\,mm blank-field survey. The most massive
  assumptions predict a halo mass that is very unlikely to be detected
  by our survey ($<$0.5\%). The least massive assumptions produce a
  mass that is well within the bounds of our survey but has a highly
  unusual gas-to-dust ratio.  The adopted mass -- dominated by
  molecular gas and dark matter -- predict that \source\ is
  87\%\ likely to be the most massive galaxy in our 2\,mm survey.  }
\label{fig:massbudget}
\end{figure}

Figure~\ref{fig:massbudget} depicts the total mass budget of the
\source\ system with different sets of assumptions drawn from
calculations in the previous sections, from the most massive set of
assumptions to the least massive set of assumptions.  All analysis
here accounts for both \source--A and \source--B together and
primarily hinge on the adopted gas and stellar masses.  The most
massive gas mass derived for \source\ uses $\alpha_{\rm
  CO}=6.5$\,\msun\,(K\,km\,s$^{-1}$\,pc$^{2}$)$^{-1}$, and the most
massive stellar mass uses that derived using {\sc Magphys}.  The least
massive set of assumptions use the more modest value of $\alpha_{\rm
  CO}=1$\,\msun\,(K\,km\,s$^{-1}$\,pc$^{2}$)$^{-1}$ and the OIR-only
derived stellar mass from \S~\ref{sec:oir}.  The dust masses are the
same for all models, as this primarily depends on $\kappa_{\rm \nu}$,
the dust mass absorption coefficient and dust temperature, for which
there is little evidence for significant dynamic range in different
environments.  The baryonic masses are then derived from the sum of
these three components.

The halo masses are then extrapolated from the total baryonic masses
using two methods, both rather conservative; these are the same
assumptions as used to estimate the halo mass of SPT0311, the most
distant DSFG \citep{marrone18a}.  The first sets an absolute floor to
the halo mass by assuming the cosmic baryon fraction $f_{\rm b}=0.19$
\citep{planck-collaboration16a} and converting from the median
baryonic mass total calculated in all cases.  This absolute lower
limit is drawn by a dotted red horizontal line in
Figure~\ref{fig:massbudget}.  The alternate technique assumes that
baryons, at most, are 5\%\ of the halo mass; in reality, this
assumption is somewhat conservative as often baryons make up less than
this, but given that this is at quite high redshift and a regime near
the tip of the mass function, 5\% is an appropriate conservative
assumption \citep[e.g.][]{behroozi18a}.  The red distributions in
Figure~\ref{fig:massbudget} show the final extrapolated halo mass
distributions, ranging from $M_{\rm
  halo}\,=\,$(6.0$\pm$1.4)$\times10^{11}$\,\msun\ (least massive
assumptions) to $M_{\rm
  halo}\,=\,$(7.2$\pm$1.8)$\times10^{12}$\,\msun\ (most massive
assumptions).

The gray regions in Figure~\ref{fig:massbudget} denote the 1$\sigma$,
2$\sigma$ and 3$\sigma$ exclusion curves for our parent survey at
$z=5.85$ calculated using the analysis described in
\citet{harrison13a}.  In other words, \source\ was selected for
follow-up out of our larger 2\,mm blank field survey covering a
contiguous area of $\sim$155\,arcmin$^{2}$, with 135\,arcmin$^{2}$ at
the necessary depth required to have identified \source\ at 5$\sigma$
significance or greater\footnote{ Note that the full
  volume of the 2\,mm map to the depth where \source\ is detectable is
  135\,arcmin$^2$, but \source\ was originally identified in a much
  smaller survey area, 9.4\,arcmin$^2$, constituting the data
  delivered in one Scheduling Block.  The rest of the 2\,mm map was
  released to us slowly over the course of the preparation of this
  manuscript.}.  The exclusion curves represent the rarest,
most-massive halos that should exist in the 135\,arcmin$^2$ survey
area at any redshift (but whose mass is evolved backward or forward to
that at $z=5.85$) with 66\%, 95\% and 99.7\%\ likelihood.  These upper
halo mass limits are 2.5$\times10^{12}$\,\msun,
4.3$\times10^{12}$\,\msun\ and 8.0$\times10^{12}$\,\msun\ for
1$\sigma$, 2$\sigma$ and 3$\sigma$, respectively.

The most massive assumptions provide an estimated halo mass that
exceeds the maximum detectable halo mass by 2.8$\sigma$; in other
words, finding a halo this massive in a survey this small is only
0.5\%\ likely.  On the other hand, the least massive assumptions
predict a total halo mass that is allowable in the given survey
limits.  Note that an intermediate solution where the lower gas mass
and higher stellar mass are adopted, the predicted halo mass is still
quite high at (4.7$\pm$1.4)$\times10^{12}$\,\msun\ by virtue of the
high stellar mass, with only a 6\%\ likelihood of having identified
such a massive system in our 2\,mm survey.  If we instead adopt the
higher gas mass and lower stellar mass, the total halo mass is
predicted to be (3.3$\pm$0.8)$\times$10$^{12}$\,\msun, with a
13\%\ likelihood of sitting in our survey volume.  
 Although it might make most sense, from a mass analysis
  standpoint, to adopt the lowest mass values, this 
would lead to
an unusual implication: the implied
  gas-to-dust ratio would be anomalously low, at GDR$\sim$16, lower
  than the vast majority of galaxies in the literature, even at
  super-solar metallicity \citep{remy-ruyer14a}. 
For this
  reason, we adopt the last intermediate mass constraints
(i.e. high gas mass, low stellar mass) due to the higher
likelihood (13\%) of occurring in the survey volume than any other
permutation.  In other words, we adopt the high gas mass using
$\alpha_{\rm CO}=6.5$\,\msun\,(K\,km\,s$^{-1}$\,pc$^{2}$)$^{-1}$, the
low stellar mass (3.2$^{+1.0}_{-1.5}$)$\times$10$^{9}$\,\msun, and
halo mass of (3.3$\pm$0.8)$\times$10$^{12}$\,\msun, as reflected in
Table~\ref{tab:physical}.

The implied gas-to-dust ratio (GDR\,=\,$M_{\rm gas}/M_{\rm dust}$) for
the system is GDR\,=\,122$^{+64}_{-43}$, in line with the canonical
ratio of 100:1 assumed for galaxies with near-solar metallicity
environments \citep{remy-ruyer14a}. Such a GDR is
  consistent with a solar-metallicity environment, though the metal
  content as not been directly constrained in \source. The gas
fraction ($f_{\rm gas}\equiv M_{\rm gas}/(M_{\rm gas}+M_\star+M_{\rm
  dust})$) is extraordinarily high at $f_{\rm
  gas}=96^{+1}_{-2}$\%\ adopting our current values; though unlike
galaxies in the nearby Universe, it is not outside of the realm of
theory to observe such gas-rich systems at $z\sim6$ or beyond.

Confirming the high gas fraction and implied halo mass require
observations of the unobscured stellar component in the mid-infrared.
These refined measurements could, in turn, lead to more
physically-motivated questions regarding the origins of galaxies like
\source.  For example, does dust at $z=5.85$ have the same absorptive
properties as local Universe dust, even if it might have a very
different origin and composition \citep[e.g.][]{de-rossi18a}? What
CO-to-H$_{\rm 2}$ conversion factor holds for \source?  

\subsection{Implications of \source\ on the prevalence of high-z DSFGs}

This source is the first that was found in a blank-field 2\,mm map
begun in ALMA Cycle 6, designed to select $z>4$ dust-obscured galaxies
efficiently.  So far, the strategy has been effective.  The initial
map in which this source was found (corresponding to one
  ALMA scheduling block) has an effective area
$\sim$9.4\,arcmin$^{2}$ with 1$\sigma_{\rm rms}\simlt0.12$\,mJy/beam
at 2\,mm\footnote{Upon later delivery of more 2\,mm data, this RMS was
  pushed deeper to the value reported in Table~\ref{tab:photometry}.},
which is the depth required to detect \source\ to 5$\sigma$
significance or above. \source\ was the only detection above this
threshold in this small map, despite several other SCUBA-2 detected
DSFGs sitting in the map region, likely indicating those other DSFGs
sit at lower redshifts $z\simlt3$.  We use the models of
\citet{casey18a} and \citet{casey18b} to comment on the possible
implications of \source\ on the number density of DSFGs at $z\simgt4$.
The first ``dust poor'' model (Model A therein) represents a universe
with very few DSFGs beyond $z>4$ and the ``dust rich'' model (Model B)
represents a universe rich with DSFGs at $z>4$.  These two models
bracket extreme interpretations of measurements in the literature, and
we refer the reader to those papers for more thorough discussion.
While the dust-poor model would predict only $\sim$0.5 sources in a
map this size and a median redshift of $z\sim3.5$ ($\pm$1$\sigma$
ranging $3<z<4.5$), the dust-rich model would predict four sources in
this map and a median redshift of $z\sim5.5$ ($\pm$1$\sigma$ ranging
$4<z<7.5$).

While a single source cannot rule either model in or out, or any
plausible model in between due to Poisson statistics, it is
interesting to note that the first source found using this mapping
strategy sits at $z=5.85$.  This is above the median redshift
predicted for both models and in the tail of the redshift distribution
predicted for the dust-poor model that is consistent with claims
from the rest-frame UV community on the lack of dust at early times.
Though tempting to infer that there might be a substantial hidden
population of DSFGs at these high redshifts based on the small survey
volume, it is also important to point out that star-formation in
massive galaxies is thought to be more heavily clustered
\citep{chiang17a} at higher and higher redshifts, such that at
$z\sim7$, half of all star-formation takes place in the progenitors of
$z=0$ galaxy clusters.  This implies that our surveys of the distant
Universe will {\it need} larger areas to not be susceptible to the
effects of cosmic variance.  Forthcoming analysis of the full 2\,mm
map dataset will provide a crucial next-step in constraining the
volume density of galaxies like \source.

\subsection{Physical Drivers of the \source\ System}

Our data suggest that \source--A and \source--B make up a close pair
of galaxies separated by 6\,kpc with a mass ratio ranging from $\sim$1:1
to $\sim$1:10 depending on the tracer.  They are likely interacting at
this close proximity, and the interaction could play a
substantial role in driving gas densities high enough to trigger
intense star formation.  At the given star-formation rates, component
A will deplete its star-forming gas in $\tau_{\rm
  depl}=38^{+16}_{-12}$\,Myr while component B will deplete its gas in
$\tau_{\rm depl}=80^{+160}_{-40}$\,Myr.  This starburst episode has
the potential to increase the stellar mass by an order of magnitude,
though the stellar mass of this system is highly uncertain.  The
star-formation rate surface densities differ an order of magnitude
between component A and B, with A consistent with some of the densest
star-forming galaxy cores known in the local Universe.

Our analysis of the halo rarity of \source\ suggests that it will live
in a massive galaxy cluster with M$_{\rm
  halo}=1.6\times10^{15}$\,\msun\ by $z=0$, as its halo is already
sufficiently massive at $z=5.85$ to constitute a node of the cosmic
web.  Though its fate is unknowable, it is likely that both components
of \source\ end up forming the very old stellar population in the core
of a brightest cluster galaxy in the Universe today.

\section{Conclusions}\label{sec:conclusions}

The ALMA data presented herein provide a uniquely detailed snapshot of
\source, the fourth-highest redshift DSFG and highest-redshift
unlensed DSFG to-date, 1\,Gyr after the Big Bang.  This system is
comprised of (at least) two galaxies that are separated by 6\,kpc
likely undergoing a merger or interaction.

The northern component, \source--A, is forming stars at nearly
$\approx$600\,\sfr\ with an incredibly dense ISM, optically thick to
the peak of dust emission near $\sim200$\,\um.  The southern source,
\source--B, is less extreme (in both star-formation surface density
and dust column density) than its northern cousin but may be of
similar mass and size.  Both components have very high attenuation in
the rest-frame UV, with $A_{\rm UV}>6.2$ and $A_{\rm
  UV}=5.0^{+1.0}_{-1.1}$ respectively. \source--A also shows some hint
at rotational motion in CO kinematics, though still potentially
dominated by dispersion (higher resolution observations are needed for
more firm kinematic diagnostics).  The measured gas depletion time for
the system is less than 100\,Myr, signaling the short-lived period of
rapid stellar growth as is often found with high-$z$ DSFGs.

The system's total halo mass is estimated to be $M_{\rm
  halo}\,=\,$(3.3$\pm$0.8)$\times10^{12}$\,\msun, which is
13\%\ likely to be the most massive galaxy detectable in our 2\,mm
blank-field survey.  This presumes that the baryonic content of
\source\ is dominated by gas (96$^{+1}_{-2}$\%) and that the stellar
mass is low, but likely to grow by an order of magnitude in a short
period of 40--80\,Myr.  Despite the multiple datasets
  providing keen insight into the physical nature of \source,
  underlying systematic uncertainties exist: the unconstrained
  CO-to-H$_{\rm 2}$ conversion factor, and the unconstrained stellar
  mass chief among them, which could significantly shift our inferred
  halo mass, gas mass, and gas depletion time.
Further analysis of the larger 2\,mm-selected sample, in addition to
future {\it JWST} constraints on the stellar emission 
  (i.e. stellar mass and metallicity) will be valuable to understand
the nature of the ISM in \source\ and its relative rarity.

We have presented a detailed fit to the long-wavelength photometry in
order to derive characteristics of dust emission at an epoch where the
CMB temperature is non-negligible; the objective of this analysis was
to illustrate the impact of different choices on the derived results,
primarily the assumed opacity of the dust.  This is most important for
systems that lack photometric constraints shortward of the dust
emission peak (at rest-frame \lpeak\,$\approx100$\,\um), which will be
{\it most} if not {\it all} galaxies identified at such epochs, with
exception of already-identified gravitationally-lensed DSFGs detected
by {\it Herschel}.  The only future telescope capable of constraining
this regime to sufficient sensitivity, providing direct insight into
the dust opacity in high-$z$ galaxies, would be NASA's future
{\it Origins Space Telescope}.

While these ALMA data have provided the basis for physical
understanding of \source, much has yet to be constrained.  Further
ALMA observations at higher spatial resolution could be used to
investigate the internal dynamics on $\simlt$kpc scales in both CO and
FIR fine-structure lines.  The VLA could constrain the full molecular
gas reservoir in low-J CO and radio continuum to provide more direct
constraints on gas mass and possible AGN, respectively.  It is
important to emphasize the role of near-future facilities like {\it
  JWST} in unlocking the stellar emission in high-$z$ sources like
\source.  As Figure~\ref{fig:fullsed} shows, the spatial resolution of
existing {\it Spitzer} data is insufficient to spatially resolve the
pair at $z=5.85$. Such systems are too faint to have been detected
by {\it Herschel}, and they might sit just at the edge of
detectability for powerful facilities like {\it HST} which is only
capable of probing rest-frame UV emission as it is heavily obscured.
{\it JWST} can not only provide a much-needed stellar mass estimate,
it can do it at much higher spatial resolution than has been possible
in the past, while it will also probe rest-frame optical nebular line
diagnostics shedding light on its metal content and strength of the
ionizing radiation field: crucial factors to measure in order to
understand the unusually dust-rich environment.

\source\ remained hidden in plain sight as one of hundreds of
850\um--1\,mm detected DSFGs without a secure redshift for years, and
even after a sensitive ALMA spectral scan and suspicions of its high
redshift, only tentative lines of marginal significance were used to
identify its redshift as $z=5.85$, first in \citet{jin19a}.  Its
identification is similar to other unlensed high-$z$ DSFGs in the
literature, in particular GN\,20 at $z=4.05$ \citep{daddi09a}, AzTEC-1
at $z=4.34$ \citep{yun15a}, and HDF\,850.1 at $z=5.18$
\citep{walter12a}, all of which took many years of effort before being
spectroscopically-confirmed by the Plateau de Bure Interferometer
(PdBI) or the Redshift Search Receiver (RSR) at the Large Millimeter
Telescope.  Though ALMA is much more sensitive than PdBI (now the
Northern Extended Millimeter Array, NOEMA) and RSR, it still requires
a time investment of a few hours on-source to provide an unequivocal
identification; such time investments per source have been rare in the
first eight years of ALMA operations.
This source, like those before it, highlights the severe need for
systematic precision source follow-up of promising high-$z$ DSFG
candidates to secure accurate constraints on the early, obscured
Universe.

\vspace{2mm}

\acknowledgements

The authors thank Shuowen Jin for helpful discussions in the
preparation of this manuscript as well as the anonymous reviewer who
provided valuable comments and suggestions.
This paper makes use of the following ALMA data: ADS/JAO.ALMA
\#2018.1.00037.A, ADS/JAO.ALMA \#2018.1.00231.S, ADS/JAO.ALMA
\#2017.1.00373.S, and ADS/JAO.ALMA \#2016.1.00279.S. ALMA is a
partnership of ESO (representing its member states), NSF (USA) and
NINS (Japan), together with NRC (Canada), MOST and ASIAA (Taiwan), and
KASI (Republic of Korea), in cooperation with the Republic of
Chile. The Joint ALMA Observatory is operated by ESO, AUI/NRAO and
NAOJ. The National Radio Astronomy Observatory is a facility of the
National Science Foundation operated under cooperative agreement by
Associated Universities, Inc.
CMC thanks the National Science Foundation for support through grants
AST-1714528 and AST-1814034, and additionally CMC and JAZ thank the
University of Texas at Austin College of Natural Sciences for support.
In addition, CMC acknowledges support from the Research Corporation
for Science Advancement from a 2019 Cottrell Scholar Award sponsored
by IF/THEN, an initiative of Lyda Hill Philanthropies.  
KIC acknowledges funding from the European Research Council through
the award of the Consolidator Grant ID 681627-BUILDUP.
SLF thanks the
National Science Foundation for support through grant AST-1518183 and
NASA through grant 80NSSC18K0954.  
KK acknowledges support from the Knut and Alice
Wallenberg Foundation. 
MT acknowledges the
support from grant PRIN MIUR 2017.
ST acknowledges support from the
ERC Consolidator Grant funding scheme (project ConTExT, grant
No. 648179).  The Cosmic DAWN Center is funded by the Danish National
Research Foundation under grant No. 140. 
 ET acknowledges support from CONICYT-Chile
grants Basal-CATA AFB-170002, FONDECYT Regular 1160999 and 1190818,
and Anillo de Ciencia y Tecnologia ACT1720033. 

\bibliography{caitlin-bibdesk}

\end{document}